\documentclass[sigconf]{acmart}
\usepackage{graphicx}
\usepackage{booktabs}
\usepackage{array}
\usepackage{enumitem}

\usepackage[inline,nomargin,index]{fixme}

\AtBeginDocument{%
  }

\copyrightyear{2026}
\acmYear{2026}
\setcopyright{cc}
\setcctype{by}
\acmConference[CHI '26]{Proceedings of the 2026 CHI Conference on Human Factors in Computing Systems}{April 13--17, 2026}{Barcelona, Spain}
\acmBooktitle{Proceedings of the 2026 CHI Conference on Human Factors in Computing Systems (CHI '26), April 13--17, 2026, Barcelona, Spain}
\acmPrice{}
\acmDOI{10.1145/3772318.3791486}
\acmISBN{979-8-4007-2278-3/2026/04}

\begin{document}

\title[Understanding Individual Cyber Resilience]{From Harm to Healing: Understanding Individual Resilience after Cybercrimes}

\author{Xiaowei Chen}
\orcid{0000-0003-0794-1551}
\affiliation{%
  \institution{Max Planck Institute for Security and Privacy}
  \city{Bochum}
  \country{Germany}
}
\email{xiaowei.chen@mpi-sp.org}

\author{Mindy Tran}
\orcid{0009-0000-1601-8588}
\affiliation{%
  \institution{Max Planck Institute for Security and Privacy}
  \city{Bochum}
  \country{Germany}
}
\email{mindy.tran@mpi-sp.org}

\author{Yue Deng}
\orcid{0009-0008-5339-8792}
\affiliation{%
  \institution{The Hong Kong University of Science and Technology}
  \city{Hong Kong}
  \country{China}
}
\email{yue.deng@mpi-sp.org}

\author{Bhupendra Acharya}
\orcid{0009-0005-3663-152X}
\affiliation{%
  \institution{University of Louisiana at Lafayette}
  \city{Louisiana}
  \country{USA}
}
\email{bhupendra.acharya@louisiana.edu}

\author{Yixin Zou}
\orcid{0000-0002-9088-705X}
\affiliation{%
  \institution{Max Planck Institute for Security and Privacy}
  \city{Bochum}
  \country{Germany}
}
\email{yixin.zou@mpi-sp.org}

\renewcommand{\shortauthors}{Chen et al.}

\begin{abstract}

How do individuals recover from cybercrimes? Victims experience various types of harm after cybercrimes, including monetary loss, data breaches, negative emotions, and even psychological trauma. The aspects that support their recovery process and contribute to individual cyber resilience remain underinvestigated. To address this gap, we interviewed 18 cybercrime victims from Western Europe using a trauma-informed approach. We identified four common stages following victimization: recognition, coping, processing, and recovery. Participants adopted various strategies to mitigate the impact of cybercrime and used different indicators to describe recovery. While they mostly relied on social support and self-regulation for emotional coping, service providers largely determined whether victims were able to recover their money. \textit{Internal factors}, \textit{external support}, and \textit{context sensitivity} collectively contribute to individuals' cyber resilience. We recommend trauma-informed support for cybercrime victims. Extending our conceptualization of individual cyber resilience, we propose collaborative and context-sensitive strategies to address the harmful impacts of cybercrime.

\end{abstract}



\maketitle

\section{Introduction}

Cybercrimes targeting individuals represent a persistent and complex global challenge. According to the Internet Crime Report 2024~\cite{fbi2024} and ENISA Threat Landscape 2024~\cite{enisa2024threat}, phishing, cryptocurrency and romance scams, malware, and ransomware were among the most reported types of cybercrime. These crimes target both organizations and individuals. In this paper, ``individuals'' refers to members of the general public. While they may be employed, we focus on their cybercrime experiences as private users, independent of workplace policies. Besides monetary loss, cybercrime victims might experience negative emotions and long-lasting psychological harms~\cite{whitty2016online,balcombe2025mental,oak2025hello}. Some victims experienced difficulties coping with post-traumatic stress, and in severe cases even suicidal thoughts~\cite{whitty2016online,balcombe2025mental}. External support may be essential in mitigating such psychological harms. For example, victim counselors can support victims to navigate this process \cite{VictimSupport2024}. However, some individuals can better cope with and adapt to the situation than others when encountering stressors~\cite{fletcher2013psychological}. A recent CHI study~\cite{wunder2025achieving} also suggested that individuals who have had prior data loss experiences were more inclined to implement data backup practices, thereby enhancing their resilience to data loss.

Cyber resilience has been a trending research topic in organizational contexts. Previous research identifies key facets of organizational cyber resilience, including resistance against potential attacks, learning from adverse attempts, and continuous adaptation to the evolving threat landscape~\cite{alhidaifi2024survey}. To enhance cyber resilience, organizations commonly implement both technical and non-technical measures, including firewalls and encryption, as well as employee training and awareness programs~\cite{alhidaifi2024survey}. Furthermore, scholars suggested that collective approaches, such as encouraging employees to report suspicious emails~\cite{lain2022phishing}, can enable swift mitigation of incoming attacks and increase an organization's cyber resilience. Many security-related decisions have been found to be context-dependent~\cite{distler_context,karyda2005information}, and prior research has examined security-related responses by focusing on specific contexts, such as employees in organizations and home computer users \cite{howe2012psychology,pfleeger2014weakest}. Nevertheless, the security knowledge and practices that employees acquire through workplace training may be transferable to their personal lives, thereby enhancing their individual cyber resilience.

To establish a clear starting point, we began our study with a preliminary definition of \textit{individual cyber resilience}, adapted from organizational contexts. We define it as ``an individual's ability to resist against cybercrimes, capacity to function continuously, and to recover from cyberattacks that target them''~\cite{alhidaifi2024survey,wunder2025achieving}. Individuals with cyber resilience can bounce back from cybercrime victimization with fewer negative residual effects on their lives~\cite{raghavan2024relationship,joinson_human_resilience}. A clear conceptual understanding of cyber resilience and related protective factors can inform a human-centered approach to create better support infrastructure for cybercrime victims~\cite{joinson_human_resilience}. While prior research has comprehensively examined the negative impacts of various cybercrimes, there has been limited investigation into the protective factors that contribute to victims’ recovery~\cite{notte2021double,balcombe2025mental}. Compared to the systematic examination of cyber resilience in organizational contexts, only a handful of studies have taken individuals' cyber resilience into their research scope~\cite{joinson_human_resilience,wunder2025achieving}. To address these gaps, we aim to investigate the following research questions (\textbf{RQ}s):

\begin{itemize}
    \item \textbf{RQ1:} How do individuals recover after experiencing cybercrimes? 
    \item \textbf{RQ2:} Which aspects support their recovery process and contribute to individual cyber resilience?
\end{itemize}

In this work, we conducted 18 trauma-informed interviews with cybercrime victims from Western Europe to understand their experiences and responses to cybercrime, identify aspects that supported their recovery, and further develop the conceptualization of individual cyber resilience. Our work makes the following \textit{contributions}:

\begin{itemize}
    \item Our conceptualization highlights how \textit{context sensitivity}, \textit{internal factors} (e.g., security knowledge and coping strategies), and \textit{external support} (e.g., social support and service providers) collectively contribute to individual cyber resilience. This extends prior understandings of individual cyber resilience and underscores the need for external supporting organizations to offer more context-sensitive and trauma-informed assistance.
    \item We reveal victims' reliance on self-regulation and social support for emotional coping, and their differing experiences with relevant stakeholders when mitigating cybercrime impacts. These results demonstrate the need for developing supporting infrastructures that are emotionally supportive and operationally responsive to victim requests.
    \item Participants recalled a range of negative emotions and frustrations after experiencing cybercrimes; this can lead to secondary traumatization of themselves as well as their first point of contact. Drawing from these insights, we made recommendations for service providers, law enforcement, and victim support organizations to foster cross-sector collaboration for addressing the harmful impacts of cybercrimes.
\end{itemize}

\section{Related Work}

First, we review the negative impacts of cybercrime on victims and the available support organizations in Section \ref{support}, which is the most closely related stream of prior literature. Then, we introduce the established research on organizational cyber resilience and discuss its positive influence on individuals' cyber resilience beyond the workplace in Section \ref{practice}. Lastly, in Section \ref{psychology} we examine relevant factors of psychological resilience, a concept well-established in psychological literature, which informs our summary of potential components of individual cyber resilience (in Table \ref{tab:resilience}).

\subsection{Impacts of Evolving Cybercrimes and External Support}
\label{support}
Individuals have long been targeted by cybercriminals. A review of nine victim surveys conducted between 2006 and 2016 in Europe revealed that online shopping fraud\footnote{We use cybercrime as a general term for illicit activities conducted via digital devices and on digital platforms. While some of the authors we reviewed refer to these activities as fraud and others as scams, we use their respective terms of choice in the Related Work section.}, online payment fraud, other types of online fraud (e.g., dating fraud), cyber threats/harassment, malware, and hacking were the most frequently surveyed types of cybercrime~\cite{junger2018victims}; however, the included types might not reflect emerging cybercrimes. A 2024 survey of U.S. residents found that 46\% of respondents reported experiencing online fraud in the past five years~\cite{acharya2024explorative}. Phishing, spoof websites, identity theft, job fraud, pig-butchering, and charity scams were among the most commonly reported types~\cite{acharya2024explorative}. Beals et al.~\cite{beals2015framework} proposed a taxonomy of fraud and identified the following seven main categories that target individuals: consumer investment fraud, consumer products and services fraud, employment fraud, prize and grant fraud, phantom debt collection fraud, charity fraud, and relationship and trust fraud. Prior work found that cybercriminals exploit various platforms and communication channels to facilitate cybercrimes targeting individuals. For instance, Acharya et al.~\cite{acharya2025pirates} analyzed a large dataset from five social media platforms and found that scammers exploit video platforms (e.g., YouTube), messaging apps (e.g., Signal, Telegram, WhatsApp), social media (e.g., X, Instagram, Facebook), and consumer platforms (e.g., Amazon, Etsy) to run charity scams.

Cybercrimes inflict various negative impacts on victims, including financial loss, compromised personal data, and psychological distress~\cite{balcombe2025mental}. The aforementioned survey with U.S. residents revealed that for individuals who encountered fraud in the past five years, the average loss amounted to \$3,209~\cite{acharya2024explorative}. Through interviews with U.S. residents, Zou et al.~\cite{zou2018ve} revealed that insufficient knowledge, costs of protective measures, optimism bias, the tendency to delay, false sense of security, and usability issues could deter individuals from proactively dealing with data breaches. Von Preuschen et al.~\cite{von2024beyond} differentiated cybersecurity-related emotions into \textit{high-arousal} and \textit{low-arousal} categories based on their level of activation; for example, frustration and annoyance (low-arousal) represent relatively mild states compared to astonishment and threat (high-arousal). Regardless of intensity, negative emotions may lead individuals to feel emotionally exhausted, hinder their productivity, and distance themselves from work~\cite{von2024beyond}. Victims of cybercrimes may develop adverse mental health symptoms. For example, a systematic review of research on cyberstalking and harassment revealed that victims commonly experienced depression, anxiety, stress, fear, and anger; furthermore, insufficient support from the criminal justice system and subsequent distrust toward technology were highlighted by Stevens et al.~\cite{stevens2021cyber}. Thus, mitigating the harms of cybercrimes necessitates measures that support victims in recovering monetary losses, safeguarding their personal data, and reducing adverse psychological effects. 

There are various victim support organizations that provide emotional support and consultation for cybercrime victims~\cite{VictimSupport2024,notte2021double}; however, the visibility and accessibility of these organizations remain unknown. Further, some UK charities that support victims of domestic abuse and stalking have struggled to address technology-facilitated abuse and have reported existing guidance getting ``out of date so quickly''~\cite{tanczer2021feel}.
Law enforcement plays a critical role in monitoring and mitigating physical crimes, but it appears to be constrained in addressing cybercrimes. Many cybercrime victims may hesitate to report their victimization to law enforcement because of stigma and negative feelings associated with being a victim; meanwhile, those who do report often receive limited assistance \cite{oak2025hello}. As an alternative, some victims turn to online platforms to discuss and validate scam cases. For instance, e-commerce seller fraud, sextortion, and corporate impostor scams are commonly discussed on Reddit \cite{scamReddit}. In general, individuals receive limited cybersecurity-related trainings, their security knowledge and practices may develop from informal sources, such as stories from friends or media outlets \cite{rader2012stories}. Furthermore, victims of cybercrimes may be unaware of the vulnerabilities exploited in their digital devices and have difficulties troubleshooting \cite{havron2019clinical}, which requires technical assistance from trained professionals.

\subsection{Organizational Cyber Resilience: Frameworks, Objectives, and Practices}
\label{practice}

Cyber resilience frameworks provide structured guidelines for organizations to identify goals, objectives, and practices addressing malicious attacks, and they reflect the possible strategies and mitigating actions that organizations can implement~\cite{bodeau2011cyber}. Bodeau et al.~\cite{bodeau2011cyber} identified four main goals for organizations: \textit{anticipate} (preparedness to adverse attacks), \textit{withstand} (ability to continue functions after attacks), \textit{recover} (restore functions after attacks), and \textit{evolve} (adapting or supporting cyber capabilities). To achieve these goals, they proposed a list of objectives to improve the systems, architectures, and functions of organizations. These objectives include ``understand, prepare, prevent, constrain, continue, reconstitute, transform, and re-architect'' \cite[p. 14]{bodeau2011cyber}. It is worth noting that the main stakeholders addressed in this framework are business heads, security officers, IT operators, system engineers, and security exercise planners. 

Organizations optimize their system engineering, architecture, and operations to increase their cyber resilience. They aimed to mitigate external attacks by enhancing IT infrastructure, operational processes, and organizational structures, with a focus on the architecture of business-critical systems~\cite{bodeau2011cyber}. In addition to these operations from business lead and security professionals, Bodeau et al.~\cite{bodeau2011cyber} noted that attackers targeted high-value (e.g., accounting department) or mission-critical resources (e.g., intellectual properties) and employees supporting those functions in adverse operations, which highlighted the critical role of employees in organizational cyber resilience. For most employees, security-related tasks are secondary to their primary job roles~\cite{brunken2023properly}. In an ideal scenario, organizations would maintain fully autonomous systems and workflows that mitigate security threats without involving employees. In practice, however, organizations depend on employees to adhere to information security policies~\cite{chen2025beyond}, participate in mandatory training~\cite{ghafur2019challenges}, practice security hygiene, and, in some cases, report suspicious activities they observed~\cite{ecabert2024implications}. Thus, Alhidaifi et al.~\cite{alhidaifi2024survey} proposed to further examine how \textit{human factors} (e.g., employee behavior and decision-making) affect organizational cyber resilience to integrate technical and non-technical measures in operations. 

Organizations implement interventions such as onboarding training, security education, and awareness programs to strengthen employees’ security practices and, in turn, enhance overall cyber resilience~\cite{odo2024strengthening}. Empirical research indicates that security education and awareness programs increase employees’ compliance intentions and security performance, improve other security-related behaviours, and reduce intentions to misuse or abuse computer systems~\cite{hu2022security}. For example, Chen et al.~\cite{chen_effects} found that in-person anti-phishing training improved employees' anti-phishing self-efficacy and support-seeking intention; as a result, they responded more securely to simulated phishing tests than the control group. With the growing adoption of `Bring Your Own Device'~\cite{ratchford2022byod}, organizations are required to take a holistic approach that engages stakeholders across different domains, including users, management, technical team, and device control. Similarly, the rise of `Work from Home' practices has extended employees’ security behaviours and hygiene practices beyond organizational premises~\cite{georgiadou2022working}. Consequently, the boundaries between personal and work devices, as well as between workplace and home security practices, have become increasingly blurred. Therefore, the security knowledge and hygiene practices employees acquire for work purposes also contribute to enhancing their individual cyber resilience. 

\subsection{Psychological Resilience: Emotion Regulation, Coping Strategy, and Trauma-informed Support}
\label{psychology}

\begin{table*}[!ht]
\centering
\caption{Potential components of individual cyber resilience informed by Related Work. * indicates emphasis by Joinson et al.~\cite{joinson_human_resilience}.}
\begin{tabular}{p{3.6cm} p{10.5cm}}
\toprule
\textbf{Sub-dimension} & \textbf{Components within Each dimension} \\
\midrule
Security Knowledge and Practices & Cybersecurity learning~\cite{ghafur2019challenges,tabassum2024drives}; self-efficacy*~\cite{joinson_human_resilience,borgert2024self}; cyber hygiene practices~\cite{vishwanath2020cyber}\\
Prior Incidents & Prior negative experience~\cite{wunder2025achieving}; Learning and growth (from incidents)*~\cite{joinson_human_resilience} \\
Problem Solving Ability & Proactively mitigate when encountering stressors, problem-focused coping~\cite{fletcher2013psychological}; hardiness~\cite{fletcher2013psychological} \\
Emotion Regulation & Experiencing positive emotion in negative circumstances~\cite{tugade2004resilient}; mindfulness, acceptance~\cite{thompson2011conceptualizing}\\
Social Support & Seek help from family, friends, and colleagues*~\cite{joinson_human_resilience}, or online forums~\cite{scamReddit}; high in extraversion~\cite{fletcher2013psychological}\\
Institutional Support & Law enforcement~\cite{batool2025between}; financial institutions~\cite{mooreshifting}; support from digital platforms (where the crime happened)~\cite{batool2025between}\\
Technical Support & Device/OS support \cite{howe2012psychology}; troubleshoot digital devices~\cite{havron2019clinical} \\
\bottomrule
\end{tabular}
\label{tab:resilience}
\end{table*}

Why do certain individuals demonstrate high resilience compared with others facing a similar stressful situation? Findings from psychological resilience provide insights. Psychological resilience can be defined as ``the ability to bounce back from negative emotional experiences and by flexible adaptation to the changing demands of stressful experiences'' \cite[p. 320]{tugade2004resilient}. This definition captures two characteristics of psychological resilience. First, resilience is closely associated with one's response to \textit{stressful incidents}. Second,  resilience can be interpreted as \textit{a positive adaptation}, to protect individuals from potential adverse impacts when dealing with stressful circumstances~\cite{fletcher2013psychological}. Furthermore, empirical studies revealed that \textit{positive emotion} supports an individual's resilience. Tugade and Fredrickson~\cite{tugade2004resilient} examined the difference in positive emotions in coping with stressful situations between low- and high-resilient individuals. They suggested that positive emotions may facilitate efficient emotion regulation, as evidenced by faster cardiovascular recovery and the derivation of positive meaning from negative circumstances. 

Psychologists have theorized resilience in terms of individual \textit{traits}, \textit{contexts}, and the \textit{processes} through which adaptive capacities are developed~\cite{raghavan2024relationship}. The conceptualization of resilience as a trait emphasizes relatively stable individual characteristics that facilitate positive adaptation to stressful situations~\cite{raghavan2024relationship}. Attributes such as extraversion and self-efficacy can therefore be understood as resilience-related traits~\cite{fletcher2013psychological}. Further, when facing different contexts, an individual may activate varied protective mechanisms; thus, examining different stressful situations allows researchers to identify relevant protective factors and to avoid overgeneralizing resilience as static traits~\cite{raghavan2024relationship}. Additionally, resilience as a process refers to research that seeks to understand how individuals achieve positive outcomes despite facing serious threats to their adaptation or development~\cite{masten2001ordinary}. Psychological resilience emerges from the dynamic interactions between personal traits and contextual factors, as well as from the transactional processes through which individuals respond to stressful situations~\cite{kuldas2022neither}.

Scholars from public health and Human-Computer Interaction advocated for trauma-informed approaches to support individuals in coping with traumatic experiences~\cite{chen2022trauma}. \textit{Trauma} can be defined as ``any disturbing experience that results in significant fear, helplessness, dissociation, confusion, or other disruptive feelings''~\cite[para. 1]{apa_trauma_definition} to a degree that it produces enduring negative effects on an individual’s attitudes and behaviors. Trauma-informed approaches emphasize four core practices: realizing the impact of trauma on individuals, recognizing its signs, responding with trauma-specific knowledge, and resisting practices that may cause re-traumatization~\cite{huang2014samhsa}. Building on SAMHSA’s framework, Chen et al.~\cite{chen2022trauma} proposed six guiding principles for \textbf{trauma-informed computing}: \textit{safety, trust, peer support, collaboration, enablement,} and \textit{intersectionality}. These principles are also relevant for cybercrime victims. They need to restore their sense of safety and trust in digital technologies exploited during the crime~\cite{stevens2021cyber}. Peer support and collaboration help victims connect with others and reduce isolation~\cite{balcombe2025mental}, while enablement emphasizes regaining control over their digital and personal lives~\cite{chen2022trauma}. Intersectionality acknowledges that victims’ experiences are shaped by their social identities and unique contexts, requiring tailored support approaches~\cite{chen2022trauma,distler_context}.

Building on the organizational and psychological resilience literature, Joinson et al.~\cite{joinson_human_resilience} developed a 14-item scale to measure ``human cyber resilience'' with four subdimensions, i.e., \textit{self-efficacy}, \textit{social support}, \textit{learning and growth}, and \textit{helplessness} (reverse-scored). While this work provides an initial measurement, Joinson et al.~\cite{joinson_human_resilience} suggested further investigation into ``individuals’ ability to recover from cyber incidents and the effectiveness of different protective strategies.'' To advance this line of research, we postulate some potential sub-dimensions and protective factors that may have been overlooked by Joinson et al.~\cite{joinson_human_resilience}. We summarize these potential components in Table \ref{tab:resilience}.

\section{Method}
We conducted semi-structured interviews due to the deeply personal nature of recovery following cybercrime victimization. This approach also helps us to capture the depth, meaning, and lived experiences of participants~\cite{carruthers1990rationale,adeoye2021research}, while minimizing intrusiveness and allowing flexibility to accommodate participants’ unique experiences~\cite{oak2025hello, whitty2016online}.

\subsection{Participants}

We refer to cybercrime victims as members of the general public who experience loss of money, personal data, digital files, time, or emotional well-being from cybercrimes~\cite{breen2022large,balcombe2025mental}. Such victimization may occur through either direct attacks targeting them or indirect attacks that affect them incidentally. To account for the unique digital infrastructure and regulatory frameworks governing online services~\cite{wicki2020budapest}, we limited participant recruitment to Western European countries. The requirements for participants were that they (1) had previously been victims of a cybercrime and experienced loss of time, money, or digital assets; (2) were fluent in English, German, or French; and (3) were at least 18 years old.

We recruited a total of 18 participants, including ten female and eight male participants residing in Western Europe. We used two approaches to recruit study participants. First, we published a recruitment questionnaire via \textit{Prolific}, which included questions on gender, age, occupation, education, the types of cybercrimes they had experienced, when the incident happened, and their willingness to be interviewed. The questionnaire recorded 180 prolific users, of whom 27 fulfilled our predefined requirements. We contacted all of them, and 11 participants (P1–P11) joined the interview sessions. Second, we also used a word-of-mouth approach to recruit cybercrime victims through our network. As a team of researchers working on cybersecurity topics, we were occasionally approached by individuals who shared their own or their family members’ cybercrime experiences. For P12-P18, we recruited them through direct contact or referrals. 

Participants' ages ranged from 22 to 65 years ($M$$=$$37.8$, $SD$$=$$11.7$). Nine participants were from the United Kingdom, and the remaining nine were from countries including Germany, France, Sweden, Luxembourg, and Denmark. Participants reported experiencing various types of cybercrime between 2018 and 2025, with most incidents occurring in 2024 ($n$$=$$5$), 2023 ($n$$=$$3$), and 2025 ($n$$=$$3$). Most participants experienced cybercrimes that directly targeted them, with only two cases involving indirect exposure—a data breach from an airline company (P1) and a family member’s phishing incident (P13). The incidents ranged from unauthorized payments and account takeovers to various types of scams, including romance, investment, rental, buyer, and task scams. Impersonation occurred in eight cases, with attackers posing as banks, crypto services, delivery companies, or trusted contacts on social platforms (see \textbf{Appendix \ref{scams}} for a more detailed account of the cybercrime experiences). With regard to education, six participants held a bachelor's degree, six a master's degree, two a PhD, three had completed high school, and one had vocational training (refer to Table \ref{participant_demo} for more details).

\begin{table*}[t]
\centering
\caption{The demographic information of participants and their experiences with cybercrimes.}
\begin{tabular}{@{}c c c p{2.5cm} p{1.5cm} p{3cm} p{1.5cm} p{1.2cm} p{1cm} p{1cm}@{}}
\toprule
\textbf{ID} & \textbf{Gender} & \textbf{Age} & \textbf{Job} & \textbf{Education} & \textbf{Cybercrime Type} & \textbf{Amount (€)} & \textbf{Loss Recovery} & \textbf{Year} & \textbf{Country} \\
\midrule
P1 & Male & 46 & Sales manager & Bachelor & Bank account compromise; data breach & 2800; Personal data & Yes/NA & 2025 & UK \\
P2 & Male & 38 & Funeral director & Vocational & Unauthorized payments & 350 & Yes & 2025 & UK \\
P3 & Female & 26 & Support worker & High school & Task scam & 580 & No & 2024 & UK \\
P4 & Female & 29 & Career counselor & Master's & Malware; Unauthorized payments & 100–200 & Yes & 2025 & UK \\
P5 & Male & 65 & Account clerk & Bachelor & Romance scam & 9000 & No & 2022 & UK \\
P6 & Female & 25 & Risk modeller & Bachelor & Malware & 550 & No & 2020 & UK \\
P7 & Male & 53 & Tram driver & High school & Impersonation (bank) & Bank info & NA & 2022 & UK \\
P8 & Male & 34 & Self-employed & Bachelor & Investment scam (crypto) & 3200 & No & 2024 & UK \\
P9 & Male & 28 & Service manager & Bachelor & Impersonation (crypto) & 220 & No & 2024 & UK \\
P10 & Female & 53 & Care assistant & High school & Buyer scam & 200 & No & 2018 & DE\\
P11 & Female & 35 & Researcher & Master's & Phishing & Email account & Yes & 2018 & DE\\
P12 & Male & 27 & Security researcher & Master's & Delivery scam & Bank info & NA & 2024 & DE\\
P13 & Female & 50+ & Strategic advisor & Master's & Phishing; impersonation (bank) & Bank info & NA & 2023 & FR \\
P14 & Female & 22 & Student & Bachelor & Delivery scam; Impersonation (bank) & 3000; Bank info & Yes/NA & 2024 & FR \\
P15 & Female & 35 & Crypto Researcher & Master's & Impersonation (WhatsApp) & Digital files & No & 2023 & LU \\
P16 & Female & 26 & Musical teacher & Master's & Rental scam & 1450 & Yes & 2023 & DK \\
P17 & Female & 43 & Material manager & PhD & Impersonation (Facebook) & FB account & Yes & 2021 & SE \\
P18 & Male & 49 & Professor & PhD & Buyer scam & 2000 & No & 2021 & SE \\
\bottomrule
\end{tabular}
\label{participant_demo}
\end{table*}

\subsection{Interview Protocol Development}

We reviewed previous interview studies on victims of deceptive chatbot scams~\cite{veisi2025user} and romance scams~\cite{whitty2016online}, as well as consumers' reactions following data breaches~\cite{zou2018ve}, prior to designing our interview protocol. The interviews centered around responding and recovering from cyberattacks, which are two key components of the individual cyber resilience definition. The final interview protocol comprised 14 questions and grouped into three parts: (a) recalling details of the cybercrime experienced by interviewees (e.g., ``Can you tell us your story of the cybercrime? Feel free to share as much as you like. This might be difficult to talk about, and you can stop whenever you want''); (b) describing how they responded to the cybercrime (e.g., ``How did you try to resolve the issue, if any?''); and (c) interviewees' reflections and recommendations regarding the cybercrime (e.g., ``What advice would you offer to others who might fall into this incident based on your experience?''). Further, the potential components of individual cyber resilience (see Table \ref{tab:resilience}), such as various stakeholders of external support, were incorporated as potential follow-up questions.

To minimize the risk of participants re-experiencing the cybercrime, the interview protocol has more questions related to recovery, lessons learned, and secure practices. In addition, two psychologists who were trained in trauma therapy reviewed the interview protocol and provided feedback prior to the interviews. We include the full interview protocol in \textbf{Appendix \ref{protocol}}. 

\subsection{Data Collection and Analysis Method}

All interview sessions were conducted remotely via Zoom. Except for P10 (in German) and P14 (in French), all other 16 interviews were conducted in English. The German and French transcripts were manually checked by native speakers before and after being translated into English. We used a GDPR-compliant, institution-licensed Copilot service to translate these two transcripts. After conducting interviews with 10 participants, we proceeded to analyze the transcripts through a structured inductive thematic analysis~\cite{braun2006using}. Furthermore, when analyzing the transcripts of the sixteenth interviewee, we observed that no new themes or codes were identified, indicating thematic saturation. To validate this observation, we interviewed two additional participants, after which we concluded data collection. We collected a total of 672 minutes of audio recordings of interviews, all of which were transcribed using the MAXQDA transcription service and thoroughly reviewed for accuracy. 

We chose an inductive thematic analysis approach and followed the guidelines suggested by Braun and Clarke~\cite{braun2021thematic}. Two authors with expertise in qualitative analysis independently developed the initial code scheme. Both of them identified meaningful text segments, generated preliminary codes, and organized these codes into category-level themes using five transcripts. The first author then integrated the independently developed two sets of categories and codes into a code scheme. Using this scheme, the first author analyzed all 18 transcripts with MAXQDA~\cite{maxqda}. The coding scheme was iteratively refined throughout the analysis by incorporating new codes as they emerged and through weekly discussions among the authors. Upon completion of coding, two authors conducted a thorough review of the analysis to ensure consistency and minimize potential misinterpretation. To present our findings, we initially grouped the coded data into three high-level themes: the impact of cybercrimes, the recovery from cybercrimes, and other themes. Furthermore, we referred to the psychological resilience literature and structured coping approaches into emotion-focused, problem-focused, and avoidant coping \cite{baker2007emotional}. Finally, we labeled the segments reflecting participants’ sense-making, adaptation, integration, and learning from the cybercrime experience as \textit{processing}. These were achieved through three iterations in manuscript drafting and multiple discussion meetings between co-authors. We include the coding scheme and exemplar quotes in the Supplementary Material.

\subsection{Ethical Considerations}

\paragraph{Trauma-informed research.} We prioritized participant well-being and ensured that distress management was embedded throughout the research process. For some individuals, retelling experiences of trauma can be empowering, whereas for others, it may be detrimental to their well-being~\cite{bolton_phases_trauma}. Inquiring about individuals' experience with cybercrime necessitates careful ethical consideration, as recalling the incident may re-traumatize them or trigger emotional distress. To mitigate these risks, we employed the following strategies: (1) The interviewer had completed training in ``Psychological First Aid''~\cite{coursera2025} and ``Trauma-informed Design Research'' (a MPI-SP workshop), which enabled them to identify and respond appropriately to signs of distress as well as to ensure self-care in conducting the research; (2) We hired an on-call psychotherapist who was available to provide emergency consulting to participants during the data collection period; and (3) Prior to the interview, we informed them that they can skip interview questions that they do not want to respond, and they can stop whenever they want. During the interviews, we chose not to prompt for more details if we observed that an interviewee did not want to share more. After the interviews, all participants were debriefed, and any questions or concerns that might have arisen from recalling their traumatic experiences were addressed. For interviewees who used self-blaming phrases, the interviewer spoke with them afterward to clarify that anyone can be targeted by cybercriminals, there is no shame in this, and the responsibility lies with the criminals.

The study was reviewed and approved by our institution’s Ethical Review Panel before data collection. We minimized the extent to which non-anonymous data were collected and stored. All audio recordings were stored on the institute's internal server and will be deleted permanently after the publication of this study. All email addresses were only kept for institutional auditing purposes. We removed all personally identifiable information from the transcripts before beginning data analysis. Participants were informed of the data we collected and their right to withdraw from the study. The median time Prolific users spent on the recruiting questionnaire was 2.32 minutes, and we compensated them with €0.5 (about €11/hr). We acquired verbal consent from the participants to audio-record the interview for transcription purposes. On average, each interview lasted 37.3 minutes ($SD$$=$$10.4$). We compensated each participant with either a €30 Prolific bonus transfer or a gift voucher.

\section{Results}

We summarize our findings descriptively to prioritize participants’ voices. Following Klemmer et al.~\cite{klemmer2025transparency}, we use quantifiers to indicate prevalence of themes among participants: ``$0\% = \text{none}$; $1$--$20\% = \text{a few}$; $21$--$40\% = \text{some}$; 
$41$--$60\% = \text{about half}$; $61$--$80\% = \text{most}$; $81$--$99\% = \text{almost all}$; 
$100\% = \text{all}$.'' 

We describe participants' cybersecurity background, situational factors, and how they identified and \textit{recognized} the cybercrimes in \ref{identify}. After recognizing the cybercrime incident, we observed three \textit{coping} approaches among our participants: emotion-focused (\ref{emotion_cop}), problem-focused (\ref{problem_cop}), and avoidant coping (\ref{avoidant}). Participants might combine different coping approaches or prioritize one approach to mitigate cybercrimes. We present how participants \textit{processed} and \textit{recovered} from cybercrime in \ref{processing} and \ref{recovery}. Lastly, we describe some gaps between current victims' needs and institutional actions in mitigating cybercrimes in \ref{gaps}.

\subsection{Recognition: The Disruption and Impact of Cybercrimes}
\label{identify}

\subsubsection{Participants' confidence, formal and informal security learning}

Overall, most participants described feeling quite confident in managing their digital device and online accounts at the moment of the interview, using terms such as ``very,'' ``pretty,'' or ``fairly'' confident. Only a few described their confidence as ``not the best,'' ``median,'' or ``low.'' Participants who described themselves as confident tended to report using VPNs and antivirus software, following password best practices, leveraging professional experience, actively managing their financial accounts, and reporting suspicious messages. In contrast, participants with lower confidence levels often described past experiences with data breaches or cyber incidents, as well as challenges with password management. A few participants mentioned having different security protections in place for their private/work and financial/social media accounts.

Participants learned their cybersecurity knowledge and practices from both formal and informal sources. For instance, some participants received structured cybersecurity training at work, while a few mentioned learning about cybersecurity-related topics through their academic studies or security-related careers (P12, P15, P18). Other common \textit{informal} learning channels were the news, TV shows, online content, and online communities, such as Reddit or YouTube. Moreover, participants highlighted the role of friends and family in shaping each other's security awareness and practices. For instance, P11 noted that their parents began using password generators after receiving advice from them, while P8 described how their grandparents encouraged them to be cautious online.

\subsubsection{Situational factors make individuals vulnerable to cybercrimes}

When participants described what they experienced during the cybercrimes, security-related knowledge and skills enabled them to recognize cyberattacks. However, situational factors, e.g., stress, distraction, and coincidental triggers, seemed to make them vulnerable to attacks. On occasions where the scammers targeted the right moment and vulnerability, it overrode their cybersecurity knowledge and skills. About half of the participants, even those who now considered themselves digitally confident, emphasized that context matters. Some participants described being deceived during moments of vulnerability, such as periods of financial need, emotional distress, anxiety, or stressful situations where they were under pressure to make quick decisions. Other participants highlighted how the scams were delivered in non-native languages, which made it difficult to critically assess their legitimacy. P12 explained: \textit{``The message was in German\dots and I am not from Germany.''} In addition, a few participants noted that their tendency to trust; a convincing message, or contextual fit and alignment, made them more likely to fall for a scam. For instance, P14 said: \textit{``It happened exactly on the day I was supposed to receive the package.''} These contextual overlaps made fraudulent messages seem credible, increasing the likelihood of deception.

\subsubsection{The realization moment: self-identification and external alerts} Some participants were able to independently identify that they had been exploited by attackers. For instance, participants' suspicion was triggered by reading device notifications, reviewing bank accounts, or reflecting on their interactions with scammers. Other participants became aware of the cybercrimes after receiving calls from their banks or warnings from close contacts. For example, P5 recalled receiving a call from their bank concerning the large transfer to a new payee, which led to the romance scam being identified and the transaction being stopped. Similarly, P15 recounted a situation where their peers warned them about a phishing message they had forwarded: ``\textit{Some [of my friends] replied to me, ‘What are you doing? What is this? Why are you sending us phishing content?’}'' Together, these examples illustrate how social validation and institutional safeguards can complement personal vigilance and lead to scam identification.

Participants with high digital confidence levels were able to promptly recognize the fraud and initiate actions to minimize its impact. In contrast, participants with lower digital confidence levels were struck with confusion and could not clearly estimate or predict the impact of the cybercrime, as P11 recalled: ``\textit{It took me two to three months to believe that really nothing happened because I thought maybe I don't see it now, but who knows, in a few weeks I’ll realize that I get huge bills from Amazon or something.}'' This difference among participants in scam identification and follow-up action illustrates the influence of security knowledge and digital confidence.

\subsubsection{Scope of Losses} Participants reported various harms resulting from cybercrimes, including financial, personal, and psychological harm. One of the most prominent themes was monetary loss. Most participants experienced financial harm ranging from €200 to €9,000. About half of the participants mentioned how exposure of their bank information, personal data, and loss of digital files caused various inconveniences for them. For instance, P13 experienced persistent phishing attempts after their personal information and bank details had been exposed: ``\textit{For six months, I received so many [phishing attempts] from Amazon, SNCF, DHL, FedEx.}'' A few participants experienced administrative harm, where they had to disrupt their schedule and productivity to resolve the scam. For instance, P2 was late for work because they had to call the bank, and P6 experienced delays in completing academic work due to the malware incident. Almost all participants reported emotional and psychological harm caused by the cybercrime incident. A few participants stopped investing in cryptocurrency due to their crypto scam experience (P6 and P8). Other participants indicated that they lost trust online or found it difficult to have conversations with people they didn’t know. Overall, the scope of losses extended beyond immediate financial damage to include psychological distress, lost time, compromised data, and even behavioral changes, illustrating the multifaceted negative impact of cybercrimes.

\subsection{Emotion-Focused Coping: Seeking Understanding Rather Than Blame} 
\label{emotion_cop}

\subsubsection{Negative emotions: from initial shock to lasting impact} Cybercrimes triggered a stream of high-arousal negative emotions, with \textit{panic} being the most frequently reported. Some participants described intense physical and mental responses upon realizing they had fallen for a scam. 
P1 shared, ``\textit{My stomach just basically did a somersault},'' while P14 said, ``\textit{I panicked, thinking he could make more transfers the next day.}'' This panic was often accompanied by \textit{stress} and \textit{anxiety}, especially among those with pre-existing medical conditions or limited time to act. P11 admitted, ``\textit{Afterwards, I still felt panicked because I thought, if they entered this email address, then gosh, what else did they enter?}'' And P16 noted, ``\textit{It was a stressful situation.}'' A few participants also expressed \textit{fear} and \textit{anger}, particularly when P13 used ``\textit{really scary}'' to describe the moment they learned that the scammer had accessed their online banking account. These immediate emotions were often compounded by confusion, in P9's words:  

\begin{quote}
 I'd say the immediate impact was mainly in terms of how I felt. Yeah. As soon as I found out the money was gone, I didn't know what to do, I was confused, I was really upset as well. I was like, what's up? What's going on? It's [sic] \textbf{confused}, \textbf{upset}, and just \textbf{worried}. (P9)   
\end{quote}

Beyond the initial shock, participants experienced a range of low-arousal negative emotions weeks or even months after the incident. \textit{Wary} and \textit{sadness} were common, especially when reflecting on financial loss or personal vulnerability. P1 stated, ``\textit{It has made me very, very wary},'' and P15 shared, ``\textit{I was just very angry and very sad for a few days}.'' \textit{Embarrassment} was reported by those who had been vigilant and confident in their security practices, as illustrated by P15, a cryptography researcher, who felt ``\textit{embarrassed because I was supposed not to be a victim. I was always vigilant and always trying to keep me, my family, and friends safe.}'' Participants also described \textit{frustration} and \textit{annoyance} due to the inconvenience and difficulty in resolving the issue, as P16 put, ``\textit{It was annoying that I couldn’t solve the problem.}'' Finally, some participants reported \textit{a hit to their ego} and lingering \textit{guilt}, as P15 reflected, ``\textit{It impacts my ego\dots I still feel guilty, deep inside me},'' and P18 added, ``\textit{It should have been clear to me\dots I shouldn’t have proceeded}.'' These enduring emotions highlighted the deep psychological toll cybercrime can have, even months after the incident.

\subsubsection{Internal and external blames} Half of our participants expressed \textit{self-blame} or \textit{blame from external parties}, indicating varied regret, self-criticism, and frustration following their victimization experiences. Some described themselves as naive or careless, and ``\textit{got myself into a situation that I shouldn't have done}'' (P1). Similarly, P3 reflected, ``\textit{I should have known that it was a scam then.}'' P5 stated, ``\textit{Although I was foolish, it was my responsibility},'' and P8 echoed this sentiment: ``\textit{It is my own fault. I can only blame myself, really. It's a lesson learned}.'' P6 described a deeper sense of blame: ``\textit{I felt a lot of self-blame and shame\dots you should have realized this is probably time to stop}.'' Participants with high digital confidence and professional expertise seemed to judge themselves more critically. For example, P18 expressed harsh self-judgment: ``\textit{How stupid could I have been\dots it should have been very apparent to me having my training.}'' These narratives reveal how cybercrime could lead to significant self-directed blame, often tied to internal factors such as perceived lapses in judgment or failure to act cautiously.

While external blame was less frequently described than self-blame, it was still present in five accounts. P16 described being blamed by their bank: ``\textit{Their [the bank’s] argument was that it was my fault because I transferred the money},'' suggesting a shift of responsibility from external parties to the victim. P6 also experienced a similar narrative when they contacted their crypto wallet: ``\textit{In their response, the blame was shifted to me because they said I gave personal, identifiable information away.}'' P6, P9, and P16 all recalled harsh remarks from their family members or friends: ``\textit{My parents were definitely disappointed in me in a certain way because\dots it's a lot of money. I think they were a bit mad at me that I just fell for this scam}'' (P16). These examples showed that while internal blame dominated participants’ reflections, external blame from banks or close ones also negatively impacted victims.

\subsubsection{Support from friends and family members} While some participants initially received ``blame'' or ``harsh comments'' from their close ones, family members, friends, and online communities were the primary sources of providing emotional support, helping victims cope with and calm negative emotions. These social actors created emotionally safe spaces where victims could process their experiences without judgment. For instance, P11 shared, ``\textit{I immediately called [my partner] and cried. But they made me rational again and told me, ‘No, don’t call me, call the phone provider, the email provider’}.'' Similarly, P18 emphasized the importance of non-judgmental support from their partner. Friends also offered emotional relief and normalization, as P10 noted, ``\textit{I talked a lot about it with my boyfriend and with a friend of mine\dots I laughed about it a bit afterwards.}'' In P9's words, ``\textit{I felt like if I was alone, it would have been a lot worse for my mental health, overthinking, just maybe depression}.'' These social interactions helped victims regain composure and begin the process of recovery.

\begin{quote}

 I had help from my dad, because he can be a little bit more like, tough. And he talked to my bank and was being a bit tougher on them. Also, he's way better at communicating and arguing with bank staff. So he was calling them as well, I was also calling them. It was a little bit stressful situation. (P16)

\end{quote}

Besides emotional support, these social actors contributed to understanding cybercrime and assisting with mitigating cybercrimes, as illustrated by the above quote from P16. Participants often turned to friends and online communities for advice and validation, as seen in P3’s experience: `\textit{`I posted [on Reddit] and a lot of people were like, yeah, this is a scam\dots That’s how I knew it was a task scam.}'' Online channels like Reddit, Telegram, and Facebook also served as collective knowledge hubs, where victims could compare their experiences with others' and confirm suspicions. As P8 shared, ``\textit{We started a group on Telegram\dots between about eight of us, we worked out what was going on.}'' In a couple of cases, family members provided technical assistance. P4 recalled, ``\textit{My brother had to help me because I didn't know how to get rid of the virus. But I remember him having to clear the laptop somehow in order to get rid of the malware}.'' These examples illustrate how social actors not only helped victims cope with negative emotions but also assisted them in addressing cyber threats.

\subsection{Problem-Focused Coping: Actions Taken to Mitigate Harm}
\label{problem_cop}

For participants who experienced malware infections or suspected that their digital devices were compromised, immediate technical remediation was a common response. Most participants experienced financial losses, which led them to seek assistance from their banks. However, participants expressed varying impressions of banks, ranging from supportive to unhelpful. Finally, because cybercriminals frequently exploited different digital platforms, some participants extended their efforts by contacting these digital platforms in search of a resolution. 

\subsubsection{Who provided technical support for cybercrime victims?} Participants with sufficient security knowledge often addressed the technical issues themselves: ``\textit{I did everything within one day, which is also something good. Being fast and effective is definitely a good sign of security handling}'' (P15). In contrast, less technically proficient participants tended to seek external technical support. P6 shared their experience of seeking technical support from their university's IT desk in the malware incident. While the IT team gave an impression of ``sarcasm'' and ``resignation,'' they successfully removed the malware and provided useful recommendations to P6. Additionally, P1 shared a unique experience in which a specialist cybersecurity team of their bank guided him remotely to reset security and scan the laptop to ``\textit{get rid of what shouldn't be there}.'' P11 recalled that the vendor who sold their parents the laptop was always helpful in installing protective software and checking whether everything was fine with the laptop.

\subsubsection{Experiences with banks vary by participant} A few participants described their banks as responsive, supportive, and efficient in handling cybercrime incidents. P1 shared a particularly positive experience in which their bank acted swiftly to secure the account, initiate an investigation, and reassure the participant that the loss was not their fault: ``\textit{They immediately took all the stress out of the situation\dots I will forever be grateful to my bank for that}.'' Similarly, P2 noted, ``\textit{My bank did a good job in how quickly it responded\dots all that money went back in 24 hours}.'' These accounts highlight the clear communication and effective actions from their banks. P13 described the contrasting responses from their two banks when they requested the usage information of two stolen cards: ``\textit{The Luxembourgish bank told me exactly at what time and place [the second day] \dots and the French bank sent me the information eight days later.}'' Some banks seemed to be more accessible and responsive when it came to supporting their clients in addressing cyber incidents.

\begin{quote}
Every time I got a representative on the phone, they’d say, ``Oh no, this isn’t the right service, you need to call someone else.'' So I’d get a new number to call, and I saw time passing. I thought, ``I’m never going to make it, it’s going to be too late\dots and that was even more frustrating because you \textbf{lose control of the situation} and have to rely on people you don’t know, not knowing whether they’ll be able to help you. (P14)  
\end{quote}

This was not a single case, as other participants encountered similar passive or constrained responses from their banks, often marked by delays, vague communication, or limited assistance. P14 described it took nearly two hours to reach the ``right'' customer service of their French bank on a Friday evening, ``\textit{during which I had no idea what was going to happen next.}'' P16 recalled being told by their Danish bank, ``\textit{Contact the police, we can't do anything},'' and felt the bank’s replies justified their inaction in tracing the money. Similarly, P18 was told by their Swedish bank that, ``\textit{it’s out of their hands}.'' These accounts suggest a lack of transparency and limited support, especially in cases involving international transfers or crypto-related fraud. P5 raised concerns about the bank’s liability, asking, ``\textit{How do they [scammers] manage to have a UK bank account without issues?}'' and suggesting that banks should work toward better fraud prevention to reduce customers’ risks. These narratives depict the victims' vulnerable situation when banks were unable or unwilling to intervene effectively.

\subsubsection{Interactions with digital platforms exploited by attackers are limited} A range of digital platforms was exploited as attack vectors in the reported cybercrimes. P2 and P11 shared their positive experience with an email service provider and eBay regarding resetting their accounts and canceling unauthorized purchases. However, a few participants attempted to contact platforms being exploited by attackers, such as crypto services and eBay, but only received generic responses. P8 commented on crypto platforms, ``\textit{They wouldn’t reply to you. They’ve just scammed everyone},'' and P10 described after reporting a scammer to eBay, P10 ``\textit{didn’t hear anything more from them}.'' By contrast, about half of the participants chose not to contact several digital platforms due to emotional barriers, distrust, or perceived ineffectiveness. P5 admitted, ``\textit{I was too embarrassed to do anything about it},'' in reference to the dating website and Instagram, and expressed skepticism about social media platforms, stating, ``\textit{They might say things, but I don’t think they actually do much}.'' P4 and P15 described that unknown outcomes and the effort required discouraged them from reaching out to British Airways or cryptocurrency websites. The perceived response efficacy, emotional readiness, required efforts, and expectations of receiving support influenced some participants' decision in reaching out to digital platforms.

\subsection{Processing: Sense-making and Learning from the Incident}
\label{processing}

\subsubsection{Rationalize, adapt, and incorporate} Participants demonstrated differing rationalization, adaptation, and integration in processing the incidents. Rationalization refers to how victims mentally process and make sense of the incident, often by evaluating their own actions, beliefs, and existing practices. P9 referenced online communities to follow suggestions and accept their loss: ``\textit{I read up on Reddit on people's other experiences\dots it made me feel, yeah, I'm not going to get that money back}.'' Some participants felt their existing security-related habits were sufficient, stating, ``\textit{the practices I was having were already good enough}'' (P12) or ``\textit{I was already doing everything I could}'' (P2). Adaptation occurred when participants took additional actions to reduce future risk and regain a sense of security. For example, P2 avoided ATMs after the card payment fraud, and P14 opened an account at a more secure bank, ``\textit{with actual advisors and proper authentication}.'' Furthermore, some participants reflect a deeper integration that goes beyond immediate reactions or short-term fixes by incorporating their victim experience into their ongoing security routines, protective tools, and decision-making processes. For example, P3 started to use website checkers for unfamiliar websites, P4 subscribed to antivirus software since then, and P13 switched to virtual cards for all online purchases. These evaluations, improvements, and habituation contributed to reconstructing a sense of security and control for participants in digital environments.

\subsubsection{More security awareness and protective practices in response to the incident}
When participants reflected on their cybercrime experience, we observed increased security awareness and vigilance following their incidents. Some participants described becoming hyper-vigilant, regularly reviewing bank statements, and being skeptical of unfamiliar messages or websites. Besides security awareness and behavioral vigilance, participants also sought additional learning resources to build on their knowledge in cybercrime topics, as P14 stated, ``\textit{Now that I’ve experienced it, I’m much more aware, so I won’t fall for it again. I’ve also watched a lot of documentaries on the topic because I got curious, I wanted to understand how these scam networks operate.}''

Following the incidents, participants adopted additional practices such as using two-factor authentication, differentiating devices for financial access, and browsing in incognito mode. Whether or not these changes successfully defended against future attacks, they provided participants with a sense of security. A few participants highlighted how this incident has prompted them to exchange security-related topics with their family and friends. For example, P15 took proactive steps to warn peers: ``\textit{
I tried to spread the word to my friends}.'' Similarly, by exchanging security incidents with family and friends, ``\textit{everybody becomes a bit more aware of what can happen}'' (P7). These reflections show that the victim's experience, while disruptive, in some cases, prompted them to adopt informed and peer-supported approaches to preventing potential cyberattacks.

\subsubsection{Advice for others}  When it came to offering advice for others who might fall victim to similar cybercrimes, participants advocated for security awareness and protective behaviors, particularly emphasizing what not to do. As P5 warned, ``\textit{Never send money to anyone you haven’t met in person},'' and P14 added, ``\textit{Never click on an SMS. Even if it means missing out on something legitimate}.'' A few participants urged users to avoid trusting social media suggestions, engaging in cryptocurrency investments, or sharing sensitive information without verification. P13 advised, ``\textit{Do not answer phone calls that you don't know},'' while P2 recommended using secure payment methods like PayPal: ``\textit{I only use PayPal for things like eBay, because I know that you've got buyer protection.}'' Similar to Geers et al.’s observations~\cite{geers2023lessons}, some participants developed resolutions that were \textit{misinformed} by their cybercrime experience. Consequently, the advice they prescribed would not directly reduce vulnerability to cyberattacks (e.g., the advice from P13 and P14 for countering impersonation). Overall, there seems to be a shared belief among participants that skepticism and caution are essential in navigating online interactions.

Furthermore, participants stressed the importance of emotional resilience, technological vigilance, and continuous learning. P1 advised, ``\textit{Keep calm and act quickly},'' while P3 reminded us, ``\textit{Don’t beat yourself up. Scams are designed to be really realistic and really smart.}'' Digital tools were seen as both a risk and a solution, as P2 noted, ``\textit{Use technology to your advantage\dots It works both ways}.'' P12 offered a sobering perspective: ``\textit{Assume that you will not be able to protect yourself just by being careful and taking actions with technology.}'' Continuous learning emerged as a recurring theme, with P14 encouraging people to ``\textit{educate yourself, there are very well-made, short documentaries explaining these practices},'' and P3 urging, ``\textit{Pass on what you learned to other people\dots it could literally save her money and save her a lot of mental health and a lot of stress.}'' Together, participants highlighted the role of emotion, technological adaptation, and learning in protecting against potential online risks.

\subsection{Recovery: Varied Indicators and Trajectories}
\label{recovery}

\subsubsection{Recovered, or not?} Participants described their recovery from cybercrime in personal and varied ways, often defining recovery not just in financial terms but also through psychological and behavioral aspects. For a few participants, recovery of monetary loss marked a clear turning point. As P14 shared, ``\textit{I was very lucky in this story because my bank reimbursed the full amount. So I didn’t have any issues on that front. For me, it was more of a lesson},'' indicating a sense of closure. P16 also noted, ``\textit{the fact that I got the money back definitely helped. If I hadn't gotten it back, I think maybe it would have felt bigger now},'' suggesting that the return of funds contributed notably to their recovery. However, some participants, despite financial recovery, expressed ongoing emotional impact, as illustrated by P1, ``\textit{I don't know that I've ever recovered from it. I'm still hyper sensitive, hyper vigilant},'' showing that recovery was not solely about money but also about regaining a sense of security.

\begin{quote}
    I think I’ve recovered. I can talk about it with people I don’t know. I still couldn’t tell my sister or friends, \textbf{but I can speak here}. I’ve probably forgotten some details because it’s been a while, and I chose not to think about it. (P5) 
\end{quote}

Some participants framed recovery in terms of emotional processing, especially when their financial losses were not recovered. P5 reflected, ``\textit{I chose not to dwell on it and tried to move on with life},'' suggesting that accepting loss and emotional distancing were key to their recovery. P6 admitted, ``\textit{It did take me a while\dots maybe two years},'' indicating a long-term recovery journey. Other participants, like P9, struggled with lingering psychological effects: ``\textit{I still get these flashbacks and struggle to sleep sometimes},'' while P7 stated, ``\textit{I don’t think you ever fully recover\dots I’ve remained very wary}.'' In contrast, a few participants described recovery as a process of increased caution and behavioral adjustment: ``\textit{I have recovered from it. I'm just very cautious. I was extremely paranoid for the first six months. I won't say I'm paranoid now, but like I said, I do not answer a phone number on my cell phone that doesn't have a name attached to it}'' (P13). These narratives show that participants attributed their recovery to different indicators, including monetary recovery, emotional calm, reconciliation, and behavioral adjustment.

\subsubsection{Key aspects supported victims' recovery} The interviews ended with a question asking about the key aspects that supported participants' recovery from the cybercrime. Some participants referred to financial institutions' timely and responsive assistance as key aspects, as P4 described: ``\textit{My bank was responsive\dots they reported it for me as fraud. So, definitely having the bank there to help.}'' Similarly, P1 and P11 shared that service providers addressed their requests professionally and reassured them that it was not their fault for being attacked. Effective responses from relevant stakeholders not only retrieved financial loss but also supported victims’ psychological recovery. As illustrated by P11:

\begin{quote}
One thing that really helped mentally was that the person from the provider who changed everything was basically just doing this, like it happens all the time. They said, ``We can’t identify the reason. \textbf{It’s not on you} that you used your name or birth date as a password.'' \textbf{That reassured me}. (P11) 
\end{quote}

Furthermore, social support played an equally vital role in recovery. Over half of our participants credited their recovery to being able to seek support from family, friends, and online communities. P6 reflected, ``\textit{The main thing was having a support network\dots I felt safe to admit my mistake and seek help.}'' Others emphasized the emotional relief of not being judged: ``\textit{No one said, ‘Oh, you’re stupid’\dots I was understood. People said, ‘Okay, it happened, now let’s find solutions’}'' (P14). Online spaces also provided knowledge and comfort, as P3 put it: ``\textit{The key aspects are people, like friends\dots not just physical friends, online communities as well.}'' These safe spaces helped victims feel less isolated and more empowered to move forward.

Being able to cope with and process cybercrime in order to reconcile with oneself was a recurring theme mentioned by participants. Some coping strategies were found to be effective, including emotion regulation, rebuilding confidence, and regaining a sense of control. Reflection and acceptance of the loss were key: ``\textit{Realizing what I had done and accepting that I had been foolish\dots I was able to move on with my life}'' (P5). Even humor helped: ``\textit{I can also laugh at myself, and I tried not to take it too seriously}'' (P10). Additionally, a few participants emphasized that viewing the incident as a learning experience contributed to their recovery: ``\textit{I just tried to put it down to a life lesson\dots Take that onwards}'' (P8). These external support and internal factors helped victims transform a distressing experience into an opportunity for developing secure practices and psychological resilience.

\subsection{Gaps in Victims' Needs and Institutional Actions}
\label{gaps}

\subsubsection{Reporting cybercrime to law enforcement: low response efficacy and disappointment}\label{police_repoorting} Most of our participants have not reported cybercrimes due to a perceived lack of efficacy and emotional barriers. As P2 pointed out that many cybercrimes were cross-jurisdictional, ``\textit{it's not really anything they [police] can do about it.}'' Others felt that the reporting to police would be futile because ``\textit{they're not going to be the ones to give me the money back}'' (P4), or believed that ``\textit{telling them that information [is] not really going to help the backlog of other crimes}'' (P9). Emotional barriers such as embarrassment and shame also played a role among a few participants; for example, P3 described, ``\textit{I felt a bit embarrassed to tell the police that I got scammed}.'' Some were deterred by the anticipated complexity of the process, with P6 noting, ``\textit{I thought it would be too much hassle. Like, it would be a lot of paperwork. Take a lot of my free time up.}''

Some participants assumed that service providers would handle the reporting automatically after they reported the cybercrimes to them. For those who did report their incident to law enforcement, the experience was often marked by a lack of follow-up or resolution. P9 shared, ``\textit{They said, ‘we’ll get back to you.’ Nobody ever got back to me\dots it was not a priority},'' while P14 recalled, ``\textit{I submitted a very detailed report\dots but I never got any follow-up on it}.'' Even when reports were acknowledged, the response was often generic and indifferent: ``\textit{They sent a letter saying, ‘We did our best, but we had to close the file’}'' (P11). These experiences reveal victims’ perceived low response efficacy and disappointment in law enforcement's responses to cybercrime.

\subsubsection{Individuals who had limited social support: avoidant coping}
\label{avoidant}

Even though social support was revealed as a key aspect that supported victims' recovery, we found that two types of participants had limited social support. First, participants with high digital literacy who acted as supporters for others received less support from their own social networks. For instance, P2, P12, and P15 specifically mentioned that they actively provided technical support to their social network regarding common attacks. Thus, they relied on their own knowledge or sought online content to mitigate the risks. This might lead to other concerns. In P2's case, they attributed their card information breach to an ATM, based on exchanges in local Facebook posts. We argue that individuals should scrutinize security resolutions from unverified sources, as platforms like Facebook, Reddit, or YouTube might not reveal the true reasons behind a data breach (P2) or serve the best interests of victims (P3). 

Second, some participants ended up using avoidant coping to mitigate the negative impacts, which reflects a tendency to avoid directly confronting or resolving the problem, instead relying on emotional numbing or psychological distancing~\cite{chao2011managing}. In P5's case, they lived by themselves and described an introverted personality: ``\textit{I’m quite introverted and tend to keep to myself.}'' They chose not to think about the romance scam as a coping strategy. Similarly, P10 and P13 demonstrated withdrawal from technology usage and a loss of trust in digital services, partly because they had not received any closure regarding the cybercrimes they experienced and thus had to move on while living with the uncertainty. Insufficient support following victimization may contribute to victims adopting avoidant coping strategies and withdrawing from digital technologies. This lack of support was also discussed by P14: ``\textit{I imagine not everyone has that kind of support. I don’t know if the government has put anything in place for the aftermath, like how to file a report, or any kind of support systems for victims of scams, or even psychological follow-ups.}'' Notably, there are \textit{victim support organizations} for cybercrimes in all of the countries where our participants reside; however, \textbf{none of them} interacted with or sought support from such organizations. 

\begin{figure*}[ht]
  \centering
  \includegraphics[width=0.95\textwidth]{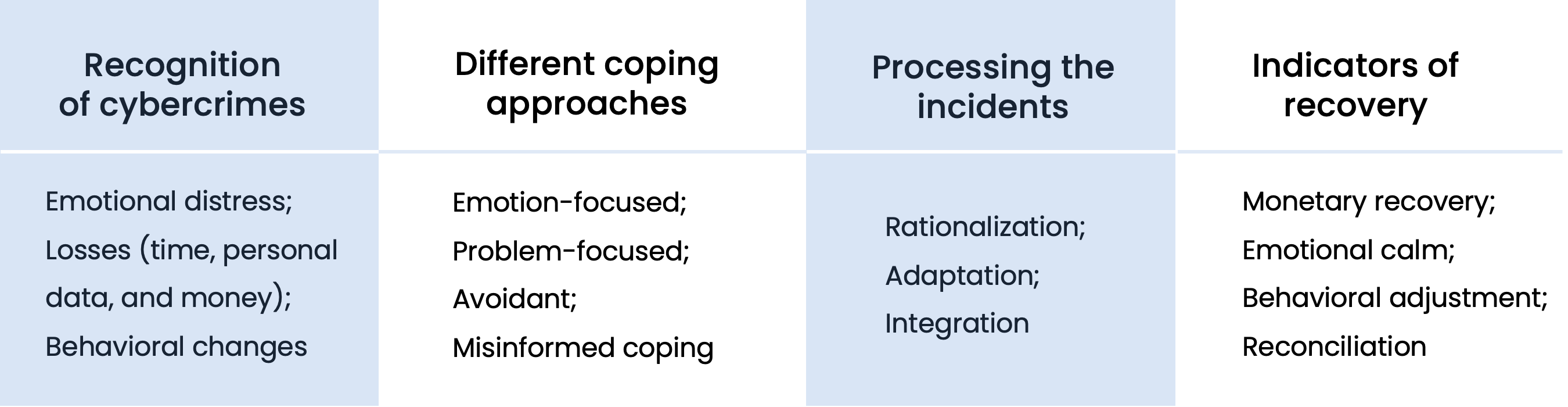}
  \caption{Visual summary: we identify four stages following cybercrime victimization in this study: recognition, coping, processing, and recovery. Note: we present these four stages thematically, without implying a chronological sequence among them.}
  \Description{Figure 1 illustrates the four stages that cybercrime victims commonly experience during their recovery: recognition, coping, processing, and recovery. When they recognize financial loss, compromise of personal data, etc., they begin to cope with the situation. Consequently, there are four coping approaches: emotion-focused, problem-focused, avoidant, and misinformed coping. Some people might move to the stage of processing, in which they rationalize, adapt, or incorporate the incident into their experience, while others might move to the recovery stage. People use different indicators to describe their recovery, for example, monetary recovery, emotional calm, behavioral adaptation, or reconciliation with oneself.}
  \label{fig-1}
\end{figure*}

\subsubsection{Platform vulnerability versus individual action}
Cybercriminals exploited a wide range of digital platforms to facilitate their attacks. Financial institutions such as banks, crypto exchanges, and money transfer companies were frequently exploited by attackers to receive payments, while e-commerce platforms were manipulated through fake profiles and fraudulent transactions. Social media platforms, including Facebook and Instagram, were commonly abused for impersonation, scammer promotion, and romance scams, and message apps like WhatsApp and Telegram were used to manipulate recipients. Food delivery services and dating platforms had weak protections for their users. Corporations such as airlines and delivery companies were commonly impersonated to target their customers and could be deceptive even for individuals with high digital security literacy. These cases demonstrate the wide vulnerability of digital platforms to being exploited by attackers to facilitate cybercrimes targeting individuals.

\begin{quote}
    Even if I got in touch with Uber Eats, I don't think I'd get anywhere. And it's not like Uber Eats is going to give me the name and address of the person who scammed my card. And then if they did, what would I do with that information? \textbf{I'm not going to hunt them down like Liam Neeson}. (P2)
\end{quote}

A few participants shared their stories about how relevant stakeholders supported them in retrieving monetary losses or mitigating the attacks. However, most participants described their disappointment with how digital platforms ``ignored'' cybercrimes that exploited their platforms. Consequently, P2 indicated that they used Uber Eats less after the unauthorized payment incident happened on the application. P17 reduced their frequency of using Facebook due to the feeling of being ``unprotected.'' P18 had not used the e-commerce website since the buyer scam on the platform, and P13 uninstalled all digital payment applications from their smartphone due to fear of leaking bank information. In P14's case, they believed their bank did a poor job in providing emergency support during the impersonation incident; thus, they switched to another bank afterwards. These cases exemplify that when victims considered that responses from digital platforms were indifferent, incompetent, or lacking care, they tended to disengage from these platforms.

\subsection{Summary of Results}

\paragraph{RQ1} Beyond monetary or digital loss, all victims went through varying degrees of psychological distress and lost time while mitigating the cybercrimes. The negative impacts of cybercrime could extend months after the incidents, including continuous attack attempts and a loss of trust and safety toward the exploited technology. Victims commonly went through four stages after encountering cybercrimes: recognition, coping, processing, and recovery (see Figure \ref{fig-1} for a visual summary). These four stages were identified thematically, without implying a chronological sequence among them. Moreover, not all victims experienced all four stages. Victims employing emotion- or problem-focused coping strategies mitigated the harms of cybercrimes more effectively than those who relied on avoidance. When moving into the processing stage, individuals rationalized what happened, adapted to the adverse incident, or integrated the experience into their daily routines. Overall, they emphasized the importance of maintaining skepticism and caution online to ensure digital safety.

\paragraph{RQ2} Individuals defined their recovery not just in terms of monetary recovery but also through emotional calm, a sense of regained control, reconciliation with themselves, and behavioral adaptation. The financial institutions' assistance to cybercrime victims was the key factor that influenced the recovery of monetary loss. Victims relied primarily on their close ones and self-regulation to cope with negative emotions, and, in some cases, close ones supplied them with technical support and problem-solving strategies. Further, being able to rebuild their confidence in managing digital devices and accounts and to draw positive lessons from their victimization experience were the key aspects that supported their recovery and developed their cyber resilience.

\section{Discussion}

\subsection{Conceptualizing Individual Cyber Resilience}

\begin{figure*}[h]
  \centering \includegraphics[width=0.8\linewidth]{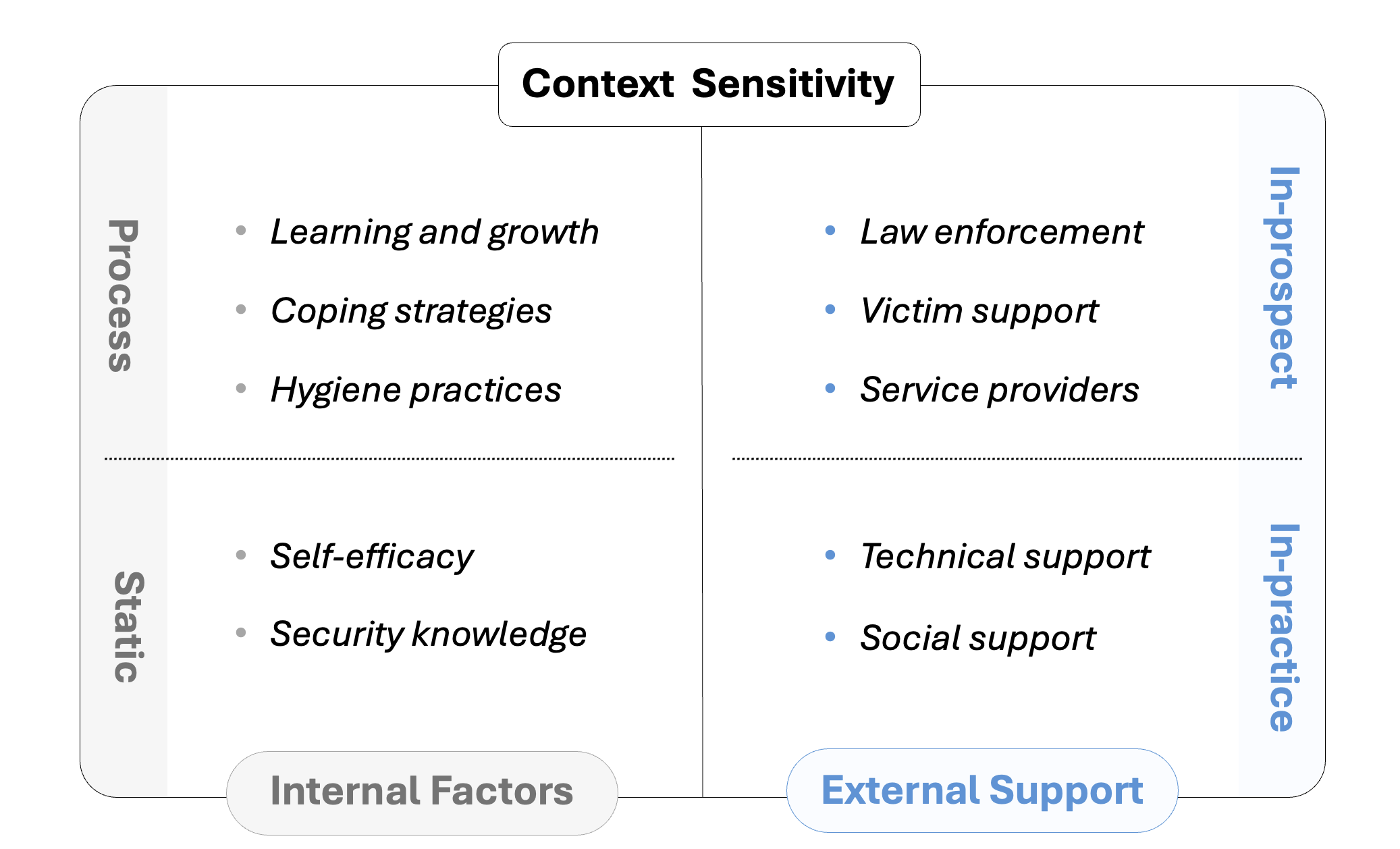}
  \caption{Visualization of individual cyber resilience.}
  \Description{Figure 2 illustrates three key dimensions that contribute to individual cyber resilience: internal factors, context sensitivity, and external support. Within internal factors, we classify security knowledge and security self-efficacy as static factors, and hygiene practices, coping strategies, and learning and growth as process factors. Within external support, we identify social support and technical support as 'in-practice,' and service providers, victim support, and law enforcement as 'in-prospect.' In the figure, we also indicate that the lines between static and process, and between 'in-practice' and 'in-prospect,' are not solid. Meanwhile, both internal factors and external support should be context sensitive.}
  \label{visual_resilience}
\end{figure*}

Several components of individual cyber resilience identified in our interviews overlap with previous conceptualizations. For example, Dupont~\cite{dupont2019cyber} described five dimensions of organizational resilience: \textit{dynamic} (encompassing activities before, during, and after incidents), \textit{networked} (the embeddedness of organizations within socio-technical systems and collaboration across units), \textit{practiced} (the outcome of sensemaking skills, surge capacity, interpersonal trust, and institutional ties), \textit{adaptive} (capacities to adapt during crises and learn from experience), and \textit{contested} (the compromises between efficiency and adaptability). We find that \textit{dynamic}, \textit{practiced}, and \textit{adaptive} are also meaningful at the individual level. In parallel, Joinson et al.~\cite{joinson_human_resilience} conceptualized \textit{human cyber resilience} with four sub-dimensions: \textit{self-efficacy} (perceived ability to respond to cyber threats), \textit{social support} (emotional and technical support from one’s networks), \textit{learning and growth} (skill development and reflection following incidents), and \textit{helplessness} (feelings that undermine resilience). Our interview findings also reflected these four sub-dimensions.

When compared with the potential components we outlined in Table \ref{tab:resilience}, some were indeed described by victims as aiding their recovery process; however, others, for example, law enforcement, victim support organizations, and service providers, were perceived as either insufficient or ineffective in victims’ recovery processes. This gap suggests that while some components are already functioning as supporting mechanisms, others require strengthening to fulfill this role effectively. Further, referring to the psychological resilience conceptualizations~\cite{kuldas2022neither}, we agree that resilience could emerge from the dynamic interaction among: (1) static traits, the protective factors individuals acquired through daily experiences; (2) processes, developed during or after exposure to adversity; and (3) context sensitivity, reflecting the ways situational cues shape protective responses. Synthesizing these perspectives, \textbf{we conceptualize individual cyber resilience with the following three dimensions}: \textit{context sensitivity}, \textit{internal factors}, and \textit{external support}. See Figure \ref{visual_resilience} for an overview of this conceptualization.

\paragraph{Context sensitivity} We define context sensitivity as individuals’ capacity to detect and adapt to situational risky cues in their technology use. Contextual factors influence individuals' susceptibility to cybercrime, such as distractions, time pressure, unfamiliar language, and social norms~\cite{distler_context,tabassum2025privacy}, or coincidental triggers. Extending this perspective to resilience, we postulate that individuals’ sensitivity to the potential risks associated with contextual factors improves their safe responses. Moreover, context is equally relevant for external supporters, which influences whether their responses and services align with victims’ lived circumstances.

\paragraph{Internal factors} Internal factors comprise resilience components that individuals develop and deploy according to their own willingness. We differentiate between static traits, which are cultivated through everyday experience, and process factors, which are activated during or after cybercrimes.

\begin{itemize}
\item \textbf{Security knowledge}: Individuals acquired security knowledge through formal and informal learning~\cite{rader2012stories}, workplace training, professional experience, and prior victimization experience. Security knowledge enhances individuals’ ability to identify unsafe interactions, recognize attacks, and anticipate potential consequences.

\item \textbf{Cybersecurity self-efficacy}: Although high self-efficacy did not eliminate susceptibility, we observed that individuals who self-reported as having high self-efficacy were able to anticipate consequences and effectively mitigate negative outcomes, largely independently of external support.

\item \textbf{Hygiene practices} safeguard the security and integrity of personal information on digital devices from cyberattacks~\cite{vishwanath2020cyber}. For instance, strong password management or multi-factor authentication provides layered defenses that reduce vulnerability to attacks, and backup practices increase resilience against data loss \cite{wunder2025achieving}. Meanwhile, we acknowledge that while maintaining a certain level of hygiene practices has been found to increase cyber resilience in both individual and organizational contexts, not everyone should become hyper-vigilant toward all types of cybercrimes. There are ``unintended harms''~\cite{chua2019identifying} and ``hidden costs''~\cite{brunken2023properly} associated with training people in these hygiene practices, and they are not the ultimate solution for countering cybercrimes.

\item \textbf{Coping strategies}: Our interviews revealed four broad categories of coping strategies after cybercrimes: emotion-focused, problem-focused, avoidant, and misinformed coping. Aligned with Baker and Berenbaum~\cite{baker2007emotional}, individuals who choose active coping strategies tend to experience more positive emotions when addressing stressors. Each coping category encompasses a range of more specific strategies that future studies could examine in-depth~\cite{jansen2018coping}.   

\item \textbf{Learning and growth}: Consistent with Joinson et al.~\cite{joinson_human_resilience}, post-incident reflection, continuous skill development, and behavioral adjustment contribute to individuals' cyber resilience. Across both individual and organizational contexts~\cite{patterson2023learning}, learning from incidents was considered one of the key preventive measures for future attacks. 
 
\end{itemize}

\paragraph{External support} External support refers to resources and interventions beyond individuals’ direct control, typically provided by other stakeholders. Further, we distinguish between components that are already ``in-practice'' and those that remain ``in-prospect.''

\begin{itemize}
\item \textbf{Social support}: Family, friends, and online forums were commonly cited as social supporters~\cite{de2020help,deng2025auntie}, and they provide emotional support, technical assistance, and problem-solving guides, helping victims both manage distress and mitigate cybercrimes.

\item \textbf{Technical support} includes performing security checks, troubleshooting compromised devices, removing malware, restoring operating system functionality, and modifying account settings to enhance security (e.g., changing passwords, enabling two-factor authentication). 

\item \textbf{Service providers} include financial institutions, telecommunication providers, and online platforms. Some financial institutions were recognized for their role in recovering monetary losses and offering timely, empathetic support. However, victims expressed reluctance to interact with online platforms that were exploited by attackers to facilitate attacks. These service providers may be in a position to identify signs of scams (e.g., in P5’s case) and malware infections \cite{bouwmeester2021thing} earlier than affected customers themselves. Accordingly, they should play a more proactive role in detecting, communicating about, and mitigating attacks that exploit their services.

\item \textbf{Law enforcement}: Victims often perceived law enforcement as having low response efficacy; however, this does not negate law enforcement’s role in enabling resilience~\cite{curtis2023understanding}. We revealed a lack of clarity from law enforcement regarding the standard procedures victims should follow after experiencing cybercrimes.

\item \textbf{Victim support} aims to provide timely emotional assistance and procedural guidance to help cybercrime victims mitigate harms~\cite{VictimSupport2024,notte2021double}. Although different victim support organizations are available to aid cybercrime victims in all the countries of our participants, none of them sought help from them. 

\end{itemize}

\subsection{Trauma-Informed Support for Cybercrime Victims}

Cybercrime victims often experience both high-arousal and low-arousal negative emotions, internal and external blame, and even long-lasting mental health impacts. Prior studies indicated that some victims even developed post-traumatic disorders and suicidal thoughts~\cite{balcombe2025mental,whitty2016online}. It is critical for service providers to provide timely and trauma-informed support for victims~\cite{meikle2024action}, because, as we observed, service providers are often the first point of contact victims reach out to mitigate cyberattacks. While cybercrime victims and survivors of technology-enabled intimate partner violence (IPV) face distinct challenges~\cite{freed2019my}, both require sensitive, trauma-informed support that addresses the technical and emotional impacts of digital harm. Following Zou et al.~\cite{zou2021role}, we recommend that service providers deploy training for customer service staff to raise awareness about the technical and emotional harms that victims experience. Furthermore, customer service agents should be equipped with trauma-informed communication strategies and guided responses to common cybercrimes, emphasizing empathy and practical assistance. More systematic examinations are needed to explore which established IPV support practices are transferable for helping cybercrime victims.

Further, some victims may recover relatively quickly, whereas those who experience emotional harm and withdraw from technologies may require more tailored support. Customer service teams should create ``respectful, welcoming, safe, and helpful'' settings and consider each victim's unique needs and the obstacles they face~\cite{elliott2005trauma}. Besides the empathy toward the primary victims of cybercrime (their customers), service staff need to caution their own vulnerability as ``second victims'' when repeatedly exposed to traumatic conversations~\cite{dekker2013second}. Informed by studies of secondary traumatization in health management~\cite{seys2013health}, customer service staff might face various negative emotional impacts in isolation, and this requires professional monitoring and intervention in place. Nevertheless, through the process of assisting victims to address incidents, it could also lead to strengthened resilience in those providing support~\cite{hernandez2007vicarious}. Organizations should implement internal supporting mechanisms to ensure that the service team receives adequate psychological counseling and emotional support, as the premise for providing empathetic victim assistance.

The blame and shame associated with cybercrime were imprinted into individuals' minds through their lived experiences, and we need to reflect on our past emphasis on shame and blame narratives in cybersecurity communication. Our participants came from diverse backgrounds, and even among cybersecurity professionals, it is indeed the case that ``anyone can be a victim.'' Prior work has shown that invoking shame in cybersecurity communication can have detrimental effects: Renaud et al.~\cite{renaud2021shame} revealed that such strategies can induce psychological distress, undermine mental well-being, disrupt personal lives, and strain workplace relationships. Likewise, we call for awareness campaigns to reduce the shame and blame associated with cybercrime victimization \cite{de2020help}. Further, given that individuals learn about cybersecurity through both formal and informal channels~\cite{wash2015too,rader2012stories}, we caution against cybersecurity communication that overemphasizes vulnerability or relies on scare tactics, as these may inadvertently exacerbate harm rather than promote psychological safety. We encourage the development of interventions that foster positive social and emotional exchange between lay people and security experts~\cite{von2025fear,gerber2025unpacking}, as well as interventions that bring intrinsic values to their target users~\cite{chen2025beyond}.



\subsection{Practical Implications and Open Challenges}

\subsubsection{Proactive consumer protection: roles of finance institutions and digital platforms} 

Some financial institutions responded to victims promptly, while others were not well-prepared to address urgent requests. In the case of UK banks, this might be partly due to the country's pioneering role in financial regulations. For example, Authorized Push Payment (APP) fraud has been subject to mandatory reimbursement rules for eligible consumers in the UK since October 7, 2024~\cite{UK2023FSMA72}. Consequently, this provision requires UK banks to share greater liability in fraud detection and real-time monitoring to counteract evolving frauds, which will result in fewer frauds exploiting bank payments~\cite{mooreshifting}. Other countries and regions should also consider adopting stricter financial regulations to protect the public. Such regulations would incentivize financial institutions to devote more resources to monitoring financial fraud and to mitigating attacks that exploit their services. Another promising direction is tailoring cyber insurance products to the needs of individual users. As an emerging topic in information security management, cyber insurance mitigates cyber risks and enhances risk management standards \cite{woods2017mapping}. Insurers should assess the cybercrime landscape targeting individuals and help consumers better understand what their policies cover \cite{hrle2025anticipating}. However, several issues, including contractual details, reporting requirements, victimization statistics and access to security solutions \cite{jain2025would}, remain unclear for individual consumers and need further investigation. 

Unfortunately, participants described that social network platforms lagged in their support and seemed not to care about widespread cybercrimes on their platforms (e.g., crypto investment scams on X, romance scams on Instagram, task scams on TikTok). These platforms, in some cases, even benefited from attackers purchasing advertisements from them~\cite{balcombe2025mental}, which were used to attract victims. Platforms should actively examine the tactics attackers use to exploit their services, implementing measures such as blocking suspicious keywords, interface warnings, and encouraging reporting~\cite{bouma2025kids}. Furthermore, more rigorous account verification should be implemented to reduce the risk of scammers creating convincing fake profiles~\cite{oak2025hello}. Social media posts often contain user metadata, link referral headers, and other information, which not only enable partial tracing of an attack’s origin but also help reconstruct the scamming chain~\cite{acharya2024imitation}. Digital platforms should leverage such metadata to help identify and block scammers on their platforms. Last but not least, individuals are almost incapable of constantly catching up with evolving cybercrimes; addressing this societal challenge requires proactive governmental intervention~\cite{renaud2018responsibilization}. The latest EU legislation introduced new rules to protect customers from fraud~\cite{eu_services}, explicitly outlining online platforms’ liability when they fail to remove fraudulent content after being informed. This provides a clear example for legislators in other regions to compel online platforms to take serious actions against cybercrimes exploiting their services.

\subsubsection{Reporting to, trainings for, and collaboration with law enforcement}

Law enforcement faces several challenges in responding effectively to support victims. Statistics indicate that most cybercrimes were not reported to authorities~\cite{freeman2024acute,sikra2023uk}. Some victims were overwhelmed by shame and embarrassment (see \ref{police_repoorting}). For those who did report, many found that law enforcement can offer limited assistance~\cite{oak2025hello}. Our study further revealed that victims are often unclear about the procedures they should follow after experiencing cybercrime, and they assumed that service providers had the duty of reporting all cybercrimes. Many victims feel cognitively burdened by the assumption to complete various forms and recall traumatic details while still emotionally affected by the incident. Experts recommend treating cybercrime victims as vulnerable individuals, offering early referrals to counseling, and ensuring empathetic and respectful handling to prevent re-victimization~\cite{whitty2016online}. Establishing victim-focused units trained in cyber psychology could improve support for cybercrime victims. To reduce cognitive demand and avoid emotional re-victimization~\cite{liu_revictimization}, an easy-to-complete reporting system is urgently needed to streamline incident documentation and facilitate victim support~\cite{agarwal2025hey,sikra2023uk}. We envision such a system as one that does not require cybercrime victims to repeatedly recall and retell their traumatic experiences whenever they interact with different stakeholders (e.g., law enforcement, financial institutions, or online platforms). Instead, the system would act as a centralized facilitator that helps victims document incidents once, notifies relevant organizations as needed, supports harm-mitigation steps, and provides tailored guidance based on the specific cybercrime encountered.  

We need to develop cross-sector initiatives in collaboration with law enforcement to address the evolving threat landscape. Only half of the surveyed Australian police officers have received cybercrime-related training, with even fewer trained in managing digital crime scenes or directing incident reports~\cite{wilson2022police}. Building the digital forensic capabilities needed to track attackers more effectively remains a significant challenge~\cite{oak2025hello}. Officers acknowledge the seriousness of cybercrime and advocate for a centralized approach that brings together multiple government agencies~\cite{cross2021responding}. However, jurisdictional issues, technological barriers, and a lack of understanding among senior officials hinder progress~\cite{curtis2023understanding,cross2021responding}. The rapidly evolving nature of cybercrime (e.g., emerging scams that combine romance manipulation with cryptocurrency investment~\cite{cross2024romance}) demands continuous training and collaboration between law enforcement, industry, and academia.


\subsubsection{More visibility and accessibility for victim support organizations} Although our study does not identify the reasons why none of the participants contacted support organizations, it may point to limited awareness or visibility of the support organizations, barriers to accessing them, or a lack of perceived relevance or trust. Limited studies have examined the interaction between cybercrime victims and support services \cite{green2020role,correia2019responding}, revealing that victim support functions not only as a procedural guide and coordination but also as emotional reassurance and prevention of counterproductive actions. Future work could explore barriers in cybercrime help-seeking to ensure that specialized support is accessible, relevant, and visible to cybercrime victims. Institutions can take several concrete steps. For instance, service providers can directly integrate victim support organizations onto their reporting portals and train first contacts of victims (e.g., customer service and police) to clearly and empathetically refer them to the relevant support organizations. Further, aligning with prior studies~\cite{bouma2024honestly}, some victims sought advice for mitigating cybercrimes from online forums. Our findings show that input from other users in online forums, such as Reddit, can also discourage users from taking proactive steps after experiencing cybercrimes. Online platforms could improve content moderation to prevent harmful advice that misguides victims or discourages them from seeking help. Future research could examine how online spaces influence victims’ coping and how moderation policies can promote more effective victim support.

\subsection{Limitations and Future Work}

In this section, we discuss some limitations of our study and propose some opportunities for future work. First, our interview protocol was carefully designed to minimize the psychological harm caused by recalling victimization experiences; however, we cannot conclude with certainty that no harm was triggered during the interviews. Future studies could incorporate trauma therapists into the study design to provide active counseling for participants. Some incidents described by participants happened several years earlier (e.g., P10 and P11). Nevertheless, they were able to recall the events with detail and consistency, for instance, the service providers involved, the impacts of the incidents, and their subsequent responses. We postulate that repeated exposure to reminders, such as using email or encountering similar online marketplaces, may have reinforced their memories over time. At the same time, we acknowledge that recalled memories can change as time passes and may be subject to recall bias. 

Second, six out of 18 participants reported intangible losses, such as stolen bank information, credentials, digital files, and other sensitive personal data. We are unable to determine the exact long-term effects of these losses, as such accounts are often harvested and later exploited by fraudsters to bypass security mechanisms. Further, as the participant sample was predominantly affected by financial scams, the emotional and psychological impacts identified may not capture the experiences of victims of other cybercrimes, such as romance scams, which can involve different and potentially more severe trauma responses.

Third, as all our participants were residents of Western Europe, the types of attacks described may be biased toward region-specific incidents. Therefore, cyberattacks targeting other regions may have been underrepresented. The authors’ language skills also constrained our choice of countries. Nevertheless, we believe the research community and practitioners can still draw insights from our findings and further examine cyber resilience in other regions. 

Fourth, our study focused on understanding cyber resilience through first-hand victim experiences. Because we did not include participants who provide external support, such as victim support organizations, financial institutions, or counter-fraud teams at digital platforms, our findings reflect only one perspective. Future work could engage different stakeholders to explore how they might proactively reach out to victims and provide timely, effective support.

Fifth, our study identified several protective factors that contribute to cyber resilience. However, it is equally important to investigate factors that hinder its development. While we focused on strategies and actions that support resilience, future work could examine behaviors, conditions, or systemic gaps that fail to support—or even reduce—individuals’ cyber resilience. Understanding these barriers would provide a more comprehensive perspective by highlighting not only what enables individuals to recover but also what obstructs or undermines that process.  

Finally, while thematic analysis is a robust method for conceptualization~\cite{naeem2023step}, grounded theory is more appropriate for building conceptual frameworks through systematic procedures~\cite{charmaz2006constructing}. Due to the sensitivity of the interview topic and the limited availability of our on-call therapist, we chose thematic analysis to address our research objectives. Future research could apply grounded theory or other methods to further refine our findings.

\section{Conclusion}

Beyond monetary loss, compromised data, and lost time, cybercrime victims experienced blame, psychological distress, loss of trust, or withdrawal from digital technologies after their incidents, even persisting for months. The protective factors and aspects that contribute to individual cyber resilience remain underinvestigated~\cite{joinson_human_resilience,balcombe2025mental}. Building upon previous work on organizational cyber resilience~\cite{dupont2019cyber,alhidaifi2024survey} and human cyber resilience~\cite{joinson_human_resilience,wunder2025achieving}, we advance the understanding of individual cyber resilience by offering an empirically grounded conceptualization. We emphasize that context sensitivity, internal factors, and external support collectively contribute to individual cyber resilience.

Through trauma-informed interviews with cybercrime victims, we reveal challenges that remain in current support infrastructures and highlight the need for context-sensitive, trauma-informed support from service providers, law enforcement, and victim support organizations. Our findings inform the design of emotionally supportive and practically equipped support infrastructures. We call on the HCI community to further explore effective reporting systems and cross-sector collaborations that address the harmful impacts of cybercrime.

\section*{Data Availability Statement}

We provide the coding scheme and exemplar quotes in the \textbf{Supplementary Material}. Due to the sensitivity of the interview scripts, we cannot share any other segments of them.

\begin{acks}

The research is partially funded by the Deutsche Forschungsgemeinschaft (DFG, German Research Foundation) under Germany’s Excellence Strategy - EXC 2092 CASA - 390781972 and a Google Academic Research Award on Trust and Safety-2024-ID 00029925. The majority of the work by Xiaowei Chen and Yue Deng was carried out while they were visiting researchers at MPI-SP. Xiaowei Chen additionally acknowledged partial support by the Young Academic Grant (2021) from the Institute for Advanced Studies at the University of Luxembourg.

\end{acks}

\bibliographystyle{ACM-Reference-Format}
\bibliography{sample-base}

@String{Computing = "Computing" }

@String{Computer = "{IEEE} Computer" }

@String{Springer = "Springer-Verlag" }

@misc{UK2023FSMA72,
  author       = {{UK Government}},
  title        = {{Financial Services and Markets Act 2023, Section 72: Liability of payment service providers for fraudulent transactions}},
  year         = {2023},
  howpublished = {\url{https://www.legislation.gov.uk/ukpga/2023/29/section/72}},
  note         = {Accessed: 2025-08-06}
}

@article{seys2013health,
  title={Health care professionals as second victims after adverse events: a systematic review},
  author={Seys, Deborah and Wu, Albert W and Gerven, Eva Van and Vleugels, Arthur and Euwema, Martin and Panella, Massimiliano and Scott, Susan D and Conway, James and Sermeus, Walter and Vanhaecht, Kris},
  journal={Evaluation \& the health professions},
  volume={36},
  number={2},
  pages={135--162},
  year={2013},
  publisher={Sage Publications Sage CA: Los Angeles, CA}
}

@article{jansen2018coping,
  title={Coping with cybercrime victimization: An exploratory study into impact and change},
  author={Jansen, Jurjen and Leukfeldt, Rutger},
  journal={Journal of Qualitative Criminal Justice and Criminology},
  volume={6},
  number={2},
  pages={205--228},
  year={2018}
}

@misc{VictimSupport2024,
  author       = {Victim Support},
  title        = {How We Can Help},
  year         = {2025},
  note         = {Accessed: 2025-08-20},
  url          = {https://www.victimsupport.org.uk/help-and-support/how-we-can-help/}
}

@article{curtis2023understanding,
  title={Understanding cybercrime in ‘real world’policing and law enforcement},
  author={Curtis, Joanna and Oxburgh, Gavin},
  journal={The Police Journal},
  volume={96},
  number={4},
  pages={573--592},
  year={2023},
  publisher={SAGE Publications Sage UK: London, England}
}

@article{renaud2018responsibilization,
  title={Is the responsibilization of the cyber security risk reasonable and judicious?},
  author={Renaud, Karen and Flowerday, Stephen and Warkentin, Merrill and Cockshott, Paul and Orgeron, Craig},
  journal={Computers \& Security},
  volume={78},
  pages={198--211},
  year={2018},
  publisher={Elsevier}
}

@article{notte2021double,
  title={Double, triple or quadruple hits? Exploring the impact of cybercrime on victims in the Netherlands},
  author={Nott{\'e}, Raoul and Leukfeldt, ER and Malsch, Marijke},
  journal={International Review of Victimology},
  volume={27},
  number={3},
  pages={272--294},
  year={2021},
  publisher={Sage Publications Sage UK: London, England}
}

@misc{coursera2025,
  author       = {George Everly},
  title        = {Psychological First Aid},
  year         = {2025},
  note         = {Coursera online course. Accessed: 2025-05-10},
  url          = {https://www.coursera.org/learn/psychological-first-aid}
}

@inproceedings{havron2019clinical,
  title={Clinical computer security for victims of intimate partner violence},
  author={Havron, Sam and Freed, Diana and Chatterjee, Rahul and McCoy, Damon and Dell, Nicola and Ristenpart, Thomas},
  booktitle={28th USENIX Security Symposium (USENIX Security 19)},
  pages={105--122},
  address={Santa Clara, CA, USA},
  publisher={USENIX Association},
  year={2019}
}

@article{correia2019responding,
  title={Responding to victimisation in a digital world: a case study of fraud and computer misuse reported in Wales},
  author={Correia, Sara Giro},
  journal={Crime Science},
  volume={8},
  number={1},
  pages={1--12},
  year={2019},
  publisher={Springer}
}

@article{green2020role,
  title={The role of victim services for individuals who have experienced serious identity-based crime},
  author={Green, Brandn and Gies, Stephen and Bobnis, Amanda and Piquero, Nicole Leeper and Piquero, Alex R and Velasquez, Eva},
  journal={Victims \& offenders},
  volume={15},
  number={6},
  pages={720--743},
  year={2020},
  publisher={Taylor \& Francis}
}

@article{sikra2023uk,
  title={UK cybercrime, victims and reporting: a systematic review},
  author={Sikra, Juraj and Renaud, Karen V and Thomas, Daniel R},
  journal={Commonwealth Cybercrime Journal},
  volume={1},
  number={1},
  pages={28--59},
  year={2023}
}

@inproceedings{geers2023lessons,
author = {Geers, Artur and Ding, Aaron and Ga\~{n}\'{a}n, Carlos Hernandez and Parkin, Simon},
title = {Lessons in Prevention and Cure: A User Study of Recovery from Flubot Smartphone Malware},
year = {2023},
isbn = {9798400708145},
publisher = {Association for Computing Machinery},
address = {New York, NY, USA},
url = {https://doi-org.proxy.bnl.lu/10.1145/3617072.3617109},
doi = {10.1145/3617072.3617109},
booktitle = {Proceedings of the 2023 European Symposium on Usable Security},
pages = {126–142},
numpages = {17},
location = {Copenhagen, Denmark},
series = {EuroUSEC '23}
}

@inproceedings{chua2019identifying,
  title={Identifying unintended harms of cybersecurity countermeasures},
  author={Chua, Yi Ting and Parkin, Simon and Edwards, Matthew and Oliveira, Daniela and Schiffner, Stefan and Tyson, Gareth and Hutchings, Alice},
  booktitle={2019 APWG Symposium on Electronic Crime Research (eCrime)},
  pages={1--15},
  year={2019},
  address={Pittsburgh, PA, USA},
  publisher={IEEE}
}

@inproceedings{von2025fear,
author = {von Preuschen, Alexandra and Benda, Carolin and Schuhmacher, Monika Christine and Zimmermann, Verena},
title = {Fear, Fun or None: A Qualitative Quest Towards Unlocking Cybersecurity Attitudes},
year = {2025},
isbn = {9798400713941},
publisher = {Association for Computing Machinery},
address = {New York, NY, USA},
url = {https://doi-org.proxy.bnl.lu/10.1145/3706598.3713538},
doi = {10.1145/3706598.3713538},
booktitle = {Proceedings of the 2025 CHI Conference on Human Factors in Computing Systems},
articleno = {1091},
numpages = {24},
location = {Yokohama, Japan},
series = {CHI '25}
}

@article{de2020help,
  title={Help, I need somebody: Examining the antecedents of social support seeking among cybercrime victims},
  author={De Kimpe, Lies and Ponnet, Koen and Walrave, Michel and Snaphaan, Thom and Pauwels, Lieven and Hardyns, Wim},
  journal={Computers in human behavior},
  volume={108},
  pages={106310},
  year={2020},
  publisher={Elsevier}
}

@inproceedings{gerber2025unpacking,
author = {Gerber, Nina and Zimmermann, Verena and Von Preuschen, Alexandra and Renaud, Karen},
title = {Unpacking the social and emotional dimensions of security and privacy user engagement},
year = {2025},
isbn = {978-1-939133-51-9},
publisher = {USENIX Association},
address = {USA},
booktitle = {Proceedings of the Twenty-First Symposium on Usable Privacy and Security},
articleno = {29},
numpages = {20},
location = {Seattle, WA, USA},
series = {SOUPS '25}
}

@inproceedings{rader2012stories,
author = {Rader, Emilee and Wash, Rick and Brooks, Brandon},
title = {Stories as informal lessons about security},
year = {2012},
isbn = {9781450315326},
publisher = {Association for Computing Machinery},
address = {New York, NY, USA},
url = {https://doi-org.proxy.bnl.lu/10.1145/2335356.2335364},
doi = {10.1145/2335356.2335364},
booktitle = {Proceedings of the Eighth Symposium on Usable Privacy and Security},
articleno = {6},
numpages = {17},
location = {Washington, D.C.},
series = {SOUPS '12}
}

@inproceedings{wash2015too,
author = {Wash, Rick and Rader, Emilee},
title = {Too much knowledge? security beliefs and protective behaviors among united states internet users},
year = {2015},
isbn = {9781931971249},
publisher = {USENIX Association},
address = {USA},
booktitle = {Proceedings of the Eleventh Symposium on Usable Privacy and Security},
pages = {309–325},
numpages = {17},
location = {Ottawa, Canada},
series = {SOUPS '15}
}

@inproceedings{zou2018ve,
author = {Zou, Yixin and Mhaidli, Abraham H. and McCall, Austin and Schaub, Florian},
title = {"I've got nothing to lose": consumers' risk perceptions and protective actions after the equifax data breach},
year = {2018},
isbn = {9781931971454},
publisher = {USENIX Association},
address = {USA},
booktitle = {Proceedings of the Fourteenth Symposium on Usable Privacy and Security},
pages = {197–216},
numpages = {20},
location = {Baltimore, MD, USA},
series = {SOUPS '18}
}

@article{bodeau2011cyber,
  title={Cyber resiliency engineering framework},
  author={Bodeau, Deborah and Graubart, Richard and Picciotto, Jeffrey and McQuaid, Rosalie},
  journal={MTR110237, MITRECorporation},
  year={2011}
}

@book{braun2021thematic,
  title={Thematic analysis: A practical guide},
  author={Braun, Virginia and Clarke, Victoria},
  year={2021},
  publisher={SAGE publications Ltd}
}

@article{acharya2024explorative,
  title={An Explorative Study of Pig Butchering Scams},
  author={Acharya, Bhupendra and Holz, Thorsten},
  journal={arXiv preprint arXiv:2412.15423},
  year={2024}
}

@inproceedings{acharya2025pirates,
author = {Acharya, Bhupendra and Lazzaro, Dario and Cin\`{a}, Antonio Emanuele and Holz, Thorsten},
title = {Pirates of Charity: Exploring Donation-based Abuses in Social Media Platforms},
year = {2025},
isbn = {9798400712746},
publisher = {Association for Computing Machinery},
address = {New York, NY, USA},
url = {https://doi-org.proxy.bnl.lu/10.1145/3696410.3714634},
doi = {10.1145/3696410.3714634},
booktitle = {Proceedings of the ACM on Web Conference 2025},
pages = {3968–3981},
numpages = {14},
location = {Sydney NSW, Australia},
series = {WWW '25}
}

@article{chao2011managing,
  title={Managing stress and maintaining well-being: Social support, problem-focused coping, and avoidant coping},
  author={Chao, Ruth Chu-Lien},
  journal={Journal of Counseling \& Development},
  volume={89},
  number={3},
  pages={338--348},
  year={2011},
  publisher={Wiley Online Library}
}

@inproceedings{breen2022large,
author = {Breen, Casey and Herley, Cormac and Redmiles, Elissa M.},
title = {A Large-Scale Measurement of Cybercrime Against Individuals},
year = {2022},
isbn = {9781450391573},
publisher = {Association for Computing Machinery},
address = {New York, NY, USA},
url = {https://doi-org.proxy.bnl.lu/10.1145/3491102.3517613},
doi = {10.1145/3491102.3517613},
booktitle = {Proceedings of the 2022 CHI Conference on Human Factors in Computing Systems},
articleno = {122},
numpages = {41},
location = {New Orleans, LA, USA},
series = {CHI '22}
}

@misc{bolton_phases_trauma,
  author       = {Mary Jo Bolton},
  title        = {Phases of Trauma Recovery},
  howpublished = {\url{https://trauma-informed.ca/recovery/phases-of-trauma-recovery/}},
  year = {n.d.},
  note         = {Accessed: 2025-08-20}
}

@misc{enisa2024threat,
  author       = {{European Union Agency for Cybersecurity (ENISA)}},
  title        = {{ENISA Threat Landscape 2024}},
  year         = {2024},
  howpublished = {\url{https://www.enisa.europa.eu/publications/enisa-threat-landscape-2024}},
  note         = {Accessed: 2025-06-03}
}

@article{raghavan2024relationship,
  title={The relationship between cultural variables and resilience to psychological trauma: A systematic review of the literature.},
  author={Raghavan, Sumithra and Sandanapitchai, Priyadharshiny},
  journal={Traumatology},
  volume={30},
  number={1},
  pages={37},
  year={2024},
  publisher={Educational Publishing Foundation}
}

@inproceedings{veisi2025user,
author = {Veisi, Omid and Kazemian, Khoshnaz and Gerami, Farzaneh and Mirzaee Kharghani, Mahya and Amirkhani, Sima and Du, Delong K. and Stevens, Gunnar and Boden, Alexander},
title = {User Narrative Study for Dealing with Deceptive Chatbot Scams Aiming to Online Fraud},
year = {2025},
isbn = {9798400713958},
publisher = {Association for Computing Machinery},
address = {New York, NY, USA},
url = {https://doi-org.proxy.bnl.lu/10.1145/3706599.3720152},
doi = {10.1145/3706599.3720152},
booktitle = {Proceedings of the Extended Abstracts of the CHI Conference on Human Factors in Computing Systems},
articleno = {560},
numpages = {7},
location = {
},
series = {CHI EA '25}
}

@article{kuldas2022neither,
  title={Neither resiliency-trait nor resilience-state: Transactional Resiliency/e},
  author={Kuldas, Seffetullah and Foody, Mair{\'e}ad},
  journal={Youth \& Society},
  volume={54},
  number={8},
  pages={1352--1376},
  year={2022},
  publisher={Sage Publications Sage CA: Los Angeles, CA}
}

@article{thompson2011conceptualizing,
  title={Conceptualizing mindfulness and acceptance as components of psychological resilience to trauma},
  author={Thompson, Rachel W and Arnkoff, Diane B and Glass, Carol R},
  journal={Trauma, Violence, \& Abuse},
  volume={12},
  number={4},
  pages={220--235},
  year={2011},
  publisher={Sage Publications Sage CA: Los Angeles, CA}
}

@article{fletcher2013psychological,
author  = {Fletcher, David and Sarkar, Mustafa},
  title   = {Psychological Resilience: A Review and Critique of Definitions, Concepts, and Theory},
  journal = {European Psychologist},
  year    = {2013},
  volume  = {18},
  number  = {1},
  pages   = {12--23},
  doi     = {10.1027/1016-9040/a000124}
}

@article{tugade2004resilient,
  title={Resilient individuals use positive emotions to bounce back from negative emotional experiences.},
  author={Tugade, Michele M and Fredrickson, Barbara L},
  journal={Journal of personality and social psychology},
  volume={86},
  number={2},
  pages={320},
  year={2004},
  publisher={American Psychological Association}
}

@article{stevens2021cyber,
  title={Cyber stalking, cyber harassment, and adult mental health: A systematic review},
  author={Stevens, Francesca and Nurse, Jason RC and Arief, Budi},
  journal={Cyberpsychology, Behavior, and Social Networking},
  volume={24},
  number={6},
  pages={367--376},
  year={2021},
  publisher={Mary Ann Liebert, Inc., publishers 140 Huguenot Street, 3rd Floor New~…}
}

@article{junger2018victims,
  author    = {Carin M. M. Reep-van den Bergh and Marianne Junger},
  title     = {Victims of cybercrime in Europe: a review of victim surveys},
  journal   = {Crime Science},
  volume    = {7},
  number    = {1},
  pages     = {5},
  year      = {2018},
  doi       = {10.1186/s40163-018-0079-3},
  url       = {https://link.springer.com/article/10.1186/s40163-018-0079-3}
}

@article{hu2022security,
  title={Security education, training, and awareness programs: Literature review},
  author={Hu, Siqi and Hsu, Carol and Zhou, Zhongyun},
  journal={Journal of Computer Information Systems},
  volume={62},
  number={4},
  pages={752--764},
  year={2022},
  publisher={Taylor \& Francis}
}

@misc{odo2024strengthening,
  title        = {Strengthening Cybersecurity Resilience: The Importance of Education, Training, and Risk Management},
  author       = {Odo, Christian},
  year         = {2024},
  month        = mar,
  howpublished = {SSRN},
  url          = {http://dx.doi.org/10.2139/ssrn.4779289}
}

@article{joinson_human_resilience,
  title={Development of a new ‘human cyber-resilience scale’},
  author={Joinson, Adam N and Dixon, Matt and Coventry, Lynne and Briggs, Pam},
  journal={Journal of Cybersecurity},
  volume={9},
  number={1},
  pages={tyad007},
  year={2023},
  publisher={Oxford University Press}
}

@inproceedings{batool2025between,
  title={Between Court Orders and Platform Policies: Understanding Law Enforcement and Meta Interactions in Addressing {Non-Consensual} Image Disclosure Abuse},
  author={Batool, Amna and Toyama, Kentaro},
  booktitle={Twenty-First Symposium on Usable Privacy and Security (SOUPS 2025)},
  publisher={USENIX Association},
  address={Seattle, WA, USA},
  pages={241--258},
  year={2025}
}

@inproceedings{borgert2024self,
author = {Borgert, Nele and Jansen, Luisa and B\"{o}se, Imke and Friedauer, Jennifer and Sasse, M. Angela and Elson, Malte},
title = {Self-Efficacy and Security Behavior: Results from a Systematic Review of Research Methods},
year = {2024},
isbn = {9798400703300},
publisher = {Association for Computing Machinery},
address = {New York, NY, USA},
url = {https://doi-org.proxy.bnl.lu/10.1145/3613904.3642432},
doi = {10.1145/3613904.3642432},
booktitle = {Proceedings of the 2024 CHI Conference on Human Factors in Computing Systems},
articleno = {973},
numpages = {32},
location = {Honolulu, HI, USA},
series = {CHI '24}
}

@inproceedings{agarwal2025hey,
  title={‘Hey mum, I dropped my phone down the toilet’: Investigating Hi Mum and Dad SMS Scams in the United Kingdom},
  author={Agarwal, Sharad and Harvey, Emma and Mariconti, Enrico and Suarez-Tangil, Guillermo and Vasek, Marie and others},
  booktitle={34th Usenix Security Symposium},
  pages={4879-4896},
  publisher={USENIX Association},
  address={Seattle, WA, USA},
  year={2025}
}

@article{patterson2023learning,
  title={Learning from cyber security incidents: A systematic review and future research agenda},
  author={Patterson, Clare M and Nurse, Jason RC and Franqueira, Virginia NL},
  journal={Computers \& Security},
  volume={132},
  pages={103309},
  year={2023},
  publisher={Elsevier}
}

@article{mooreshifting,
  title={How Shifting Liability Explains Rising Cybercrime Costs},
  author={Moore, Tyler},
  journal={Rossfest Festschrift},
  year={2024},
  pages={135}
}

@article{freeman2024acute,
  title={Acute and diffuse impacts of fraud: A victim-centred teleology for a wicked problem},
  author={Freeman, Christopher},
  journal={Journal of Economic Criminology},
  volume={6},
  pages={100104},
  year={2024},
  publisher={Elsevier}
}

@inproceedings{tabassum2024drives,
author = {Tabassum, Sarah and Faklaris, Cori and Lipford, Heather Richter},
title = {What drives SMiShing susceptibility? a U.S. interview study of how and why mobile phone users judge text messages to be real or fake},
year = {2024},
isbn = {978-1-939133-42-7},
publisher = {USENIX Association},
address = {USA},
booktitle = {Proceedings of the Twentieth Symposium on Usable Privacy and Security},
articleno = {21},
numpages = {19},
location = {Philadelphia, PA, USA},
series = {SOUPS '24}
}

@inproceedings{tabassum2025privacy,
author = {Tabassum, Sarah and Mathew, Nishka and Faklaris, Cori},
title = {Privacy on the Move: Understanding Educational Migrants’ Social Media Practices through the Lens of Communication Privacy Management Theory},
year = {2025},
isbn = {9798400714849},
publisher = {Association for Computing Machinery},
address = {New York, NY, USA},
url = {https://doi-org.proxy.bnl.lu/10.1145/3715335.3735453},
doi = {10.1145/3715335.3735453},
booktitle = {Proceedings of the 2025 ACM SIGCAS/SIGCHI Conference on Computing and Sustainable Societies},
pages = {1–18},
numpages = {18},
location = {
},
series = {COMPASS '25}
}

@inproceedings{bouma2024honestly,
author = {Bouma-Sims, Elijah and Hassan, Hiba and Nisenoff, Alexandra and Cranor, Lorrie Faith and Christin, Nicolas},
title = {"It was honestly just gambling": investigating the experiences of teenage cryptocurrency users on reddit},
year = {2024},
isbn = {978-1-939133-42-7},
publisher = {USENIX Association},
address = {USA},
booktitle = {Proceedings of the Twentieth Symposium on Usable Privacy and Security},
articleno = {18},
numpages = {20},
location = {Philadelphia, PA, USA},
series = {SOUPS '24}
}

@inproceedings{bouma2025kids,
  title={The Kids Are All Right: Investigating the Susceptibility of Teens and Adults to YouTube Giveaway Scams.},
  author={Bouma-Sims, Elijah Robert and Klucinec, Lily and Lanyon, Mandy and Downs, Julie and Cranor, Lorrie Faith},
  booktitle={Network and Distributed System Security Symposium 2025},
  publisher={NDSS},
  address={San Diego, CA, USA},
  pages={1-18},
  year={2025}
}

@article{meikle2024action,
  title={“What action should l take?”: Help-seeking behaviours of those targeted by romance fraud},
  author={Meikle, Wesley and Cross, Cassandra},
  journal={Journal of Economic Criminology},
  volume={3},
  pages={100054},
  year={2024},
  publisher={Elsevier}
}

@article{wicki2020budapest,
  title={The Budapest Convention and the General Data Protection Regulation: acting in concert to curb cybercrime?},
  author={Wicki-Birchler, David},
  journal={International Cybersecurity Law Review},
  volume={1},
  number={1},
  pages={63--72},
  year={2020},
  publisher={Springer}
}

@inproceedings{zou2021role,
  title={The role of computer security customer support in helping survivors of intimate partner violence},
  author={Zou, Yixin and McDonald, Allison and Narakornpichit, Julia and Dell, Nicola and Ristenpart, Thomas and Roundy, Kevin and Schaub, Florian and Tamersoy, Acar},
  booktitle={30th USENIX security Symposium (USENIX Security 21)},
  address={virtual event},
  publisher={USENIX Association},
  pages={429--446},
  year={2021}
}

@article{wilson2022police,
  title={Police preparedness to respond to cybercrime in Australia: An analysis of individual and organizational capabilities},
  author={Wilson, Michael and Cross, Cassandra and Holt, Thomas and Powell, Anastasia},
  journal={Journal of Criminology},
  volume={55},
  number={4},
  pages={468--494},
  year={2022},
  publisher={Sage Publications Sage UK: London, England}
}

@inproceedings{klemmer2025transparency,
  title={Transparency in Usable Privacy and Security Research: Scholars' Perspectives, Practices, and Recommendations},
  author={Klemmer, Jan H and Schm{\"u}ser, Juliane and Lowens, Byron M and Fischer, Fabian and Schm{\"u}ser, Lea and Schaub, Florian and Fahl, Sascha},
  booktitle={2025 IEEE Symposium on Security and Privacy (SP)},
  pages={2658--2677},
  address={San Francisco, CA, USA},
  year={2025},
  publisher={IEEE}
}

@book{cross2021responding,
  title={Responding to cybercrime: Perceptions and need of Australian police and the general community},
  author={Cross, Cassandra and Holt, Thomas and Powell, Anastasia and Wilson, Michael},
  year={2021},
  address={Canberra, Australia},
  publisher={Australian Institute of Criminology}
}

@article{cross2024romance,
  title={Romance baiting, cryptorom and ‘pig butchering’: an evolutionary step in romance fraud},
  author={Cross, Cassandra},
  journal={Current Issues in Criminal Justice},
  volume={36},
  number={3},
  pages={334--346},
  year={2024},
  publisher={Taylor \& Francis}
}

@article{carruthers1990rationale,
  title={A Rationale for the Use of Semi-structured Interviews},
  author={Carruthers, John},
  journal={Journal of Educational Administration},
  volume={28},
  number={1},
  year={1990},
  publisher={MCB UP Ltd}
}

@article{adeoye2021research,
  title={Research and scholarly methods: Semi-structured interviews},
  author={Adeoye-Olatunde, Omolola A and Olenik, Nicole L},
  journal={Journal of the american college of clinical pharmacy},
  volume={4},
  number={10},
  pages={1358--1367},
  year={2021},
  publisher={Wiley Online Library}
}

@article{hernandez2007vicarious,
  title={Vicarious resilience: A new concept in work with those who survive trauma},
  author={Hern{\'a}ndez, Pilar and Gangsei, David and Engstrom, David},
  journal={Family process},
  volume={46},
  number={2},
  pages={229--241},
  year={2007},
  publisher={Wiley Online Library}
}

@inproceedings{acharya2024imitation,
  title={The imitation game: Exploring brand impersonation attacks on social media platforms},
  author={Acharya, Bhupendra and Lazzaro, Dario and L{\'o}pez-Morales, Efr{\'e}n and Oest, Adam and Saad, Muhammad and Cin{\`a}, Antonio Emanuele and Sch{\"o}nherr, Lea and Holz, Thorsten},
  booktitle={33rd USENIX Security Symposium (USENIX Security 24)},
  publisher={USENIX Association},
  address={Philadelphia, PA, USA},
  pages={4427--4444},
  year={2024}
}

@article{whitty2016online,
  title={The online dating romance scam: The psychological impact on victims--both financial and non-financial},
  author={Whitty, Monica T and Buchanan, Tom},
  journal={Criminology \& Criminal Justice},
  volume={16},
  number={2},
  pages={176--194},
  year={2016},
  publisher={Sage Publications Sage UK: London, England}
}

@misc{maxqda,
  title={MAXQDA},
  author={VERBIsoftware},
  year={2024},
  howpublished = {\url{https://www.maxqda.com/}},
  note = {accessed: 2025-08-10}
}

@inproceedings{liu_revictimization,
author = {Liu, Peiyao and Su, Norman Makoto},
title = {Emotional Re-Victimization in the Workplace: The Burden of Concern Reporting Systems},
year = {2025},
isbn = {9798400713842},
publisher = {Association for Computing Machinery},
address = {New York, NY, USA},
url = {https://doi.org/10.1145/3729176.3729188},
doi = {10.1145/3729176.3729188},
booktitle = {Proceedings of the 4th Annual Symposium on Human-Computer Interaction for Work},
articleno = {18},
numpages = {13},
location = {
},
series = {CHIWORK '25}
}

@book{dekker2013second,
  title={Second victim: Error, guilt, trauma, and resilience},
  author={Dekker, Sidney},
  year={2013},
  address={Boca Raton, USA},
  publisher={CRC press}
}

@inproceedings{renaud2021shame,
author = {Renaud, Karen and Searle, Rosalind and Dupuis, Marc},
title = {Shame in Cyber Security: Effective Behavior Modification Tool or Counterproductive Foil?},
year = {2022},
isbn = {9781450385732},
publisher = {Association for Computing Machinery},
address = {New York, NY, USA},
url = {https://doi-org.proxy.bnl.lu/10.1145/3498891.3498896},
doi = {10.1145/3498891.3498896},
booktitle = {Proceedings of the 2021 New Security Paradigms Workshop},
pages = {70–87},
numpages = {18},
location = {Virtual Event, USA},
series = {NSPW '21}
}

@article{dupont2019cyber,
  title={The cyber-resilience of financial institutions: significance and applicability},
  author={Dupont, Beno{\^\i}t},
  journal={Journal of cybersecurity},
  volume={5},
  number={1},
  pages={tyz013},
  year={2019},
  publisher={Oxford University Press}
}

@article{braun2006using,
  title={Using thematic analysis in psychology},
  author={Braun, Virginia and Clarke, Victoria},
  journal={Qualitative research in psychology},
  volume={3},
  number={2},
  pages={77--101},
  year={2006},
  publisher={Taylor \& Francis}
}

@article{elliott2005trauma,
  title={Trauma-informed or trauma-denied: Principles and implementation of trauma-informed services for women},
  author={Elliott, Denise E and Bjelajac, Paula and Fallot, Roger D and Markoff, Laurie S and Reed, Beth Glover},
  journal={Journal of community psychology},
  volume={33},
  number={4},
  pages={461--477},
  year={2005},
  publisher={Wiley Online Library}
}

@article{balcombe2025mental,
  title={The Mental Health Impacts of Internet Scams},
  author={Balcombe, Luke},
  journal={International Journal of Environmental Research and Public Health},
  volume={22},
  number={6},
  pages={938},
  year={2025},
  publisher={MDPI}
}

@article{masten2001ordinary,
  title={Ordinary magic: Resilience processes in development.},
  author={Masten, Ann S},
  journal={American psychologist},
  volume={56},
  number={3},
  pages={227},
  year={2001},
  publisher={American Psychological Association}
}

@inproceedings{chen_effects,
author = {Chen, Xiaowei and Sacr\'{e}, Margault and Lenzini, Gabriele and Greiff, Samuel and Distler, Verena and Sergeeva, Anastasia},
title = {The Effects of Group Discussion and Role-playing Training on Self-efficacy, Support-seeking, and Reporting Phishing Emails: Evidence from a Mixed-design Experiment},
year = {2024},
isbn = {9798400703300},
publisher = {Association for Computing Machinery},
address = {New York, NY, USA},
url = {https://doi.org/10.1145/3613904.3641943},
doi = {10.1145/3613904.3641943},
booktitle = {Proceedings of the 2024 CHI Conference on Human Factors in Computing Systems},
articleno = {829},
numpages = {21},
location = {Honolulu, HI, USA},
series = {CHI '24}
}

@article{georgiadou2022working,
  title={Working from home during COVID-19 crisis: a cyber security culture assessment survey},
  author={Georgiadou, Anna and Mouzakitis, Spiros and Askounis, Dimitris},
  journal={Security Journal},
  volume={35},
  number={2},
  pages={486--505},
  year={2022},
  publisher={Springer}
}

@article{ratchford2022byod,
  title={BYOD security issues: A systematic literature review},
  author={Ratchford, Melva and El-Gayar, Omar and Noteboom, Cherie and Wang, Yong},
  journal={Information Security Journal: A Global Perspective},
  volume={31},
  number={3},
  pages={253--273},
  year={2022},
  publisher={Taylor \& Francis}
}

@inproceedings{distler_context,
author = {Distler, Verena},
title = {The Influence of Context on Response to Spear-Phishing Attacks: an In-Situ Deception Study},
year = {2023},
isbn = {9781450394215},
publisher = {Association for Computing Machinery},
address = {New York, NY, USA},
url = {https://doi.org/10.1145/3544548.3581170},
doi = {10.1145/3544548.3581170},
booktitle = {Proceedings of the 2023 CHI Conference on Human Factors in Computing Systems},
articleno = {619},
numpages = {18},
location = {Hamburg, Germany},
series = {CHI '23}
}

@article{huang2014samhsa,
  title={SAMHSA's concept of truama and guidance for a trauma-informed approach},
  author={Huang, Larke N and Flatow, Rebecca and Biggs, Tenly and Afayee, Sara and Smith, Kelley and Clark, Thomas and Blake, Mary},
  year={2014},
  publisher={Substance Abuse and Mental Health Services Administration (SAMHSA)}
}

@misc{apa_trauma_definition,
  title        = {“Trauma”},
  author       = {{American Psychological Association}},
  year         = {n.d.},
  howpublished = {\emph{APA Dictionary of Psychology}},
  url          = {https://dictionary.apa.org/trauma},
  note         = {Accessed: 2025-08-28}
}

@inproceedings{chen2022trauma,
author = {Chen, Janet X. and McDonald, Allison and Zou, Yixin and Tseng, Emily and Roundy, Kevin A and Tamersoy, Acar and Schaub, Florian and Ristenpart, Thomas and Dell, Nicola},
title = {Trauma-Informed Computing: Towards Safer Technology Experiences for All},
year = {2022},
isbn = {9781450391573},
publisher = {Association for Computing Machinery},
address = {New York, NY, USA},
url = {https://doi-org.proxy.bnl.lu/10.1145/3491102.3517475},
doi = {10.1145/3491102.3517475},
booktitle = {Proceedings of the 2022 CHI Conference on Human Factors in Computing Systems},
articleno = {544},
numpages = {20},
location = {New Orleans, LA, USA},
series = {CHI '22}
}

@article{woods2017mapping,
  title={Mapping the coverage of security controls in cyber insurance proposal forms},
  author={Woods, Daniel and Agrafiotis, Ioannis and Nurse, Jason RC and Creese, Sadie},
  journal={Journal of Internet Services and Applications},
  volume={8},
  number={1},
  pages={8},
  year={2017},
  publisher={Springer}
}

@inproceedings{jain2025would,
  title={“Why would money protect me from cyber bullying?”: A mixed-methods study of personal cyber insurance},
  author={Jain, Rachiyta and Hrle, Temima and Marinetti, Margherita and Jenkins, Adam and B{\"o}hme, Rainer and Woods, Daniel W},
  booktitle={2025 IEEE Symposium on Security and Privacy (SP)},
  pages={2264--2283},
  year={2025},
  address={San Francisco, CA, USA},
  publisher={IEEE}
}

@inproceedings{hrle2025anticipating,
  title={Anticipating Personal Cyber Insurance Disputes: A US/UK User Study},
  author={Hrle, Temima and Piao, Yangheran and Woods, Daniel},
  booktitle={The Workshop on the Economics of Information Security},
  address={Tokyo, Japan},
  pages={1-22},
  publisher={WEIS2025},
  year={2025}
}

@inproceedings{bouwmeester2021thing,
author = {Bouwmeester, Brennen and Rodr\'{\i}guez, Elsa and Ga\~{n}\'{a}n, Carlos and Van Eeten, Michel and Parkin, Simon},
title = {"The thing doesn't have a name": learning from emergent real-world interventions in smart home security},
year = {2021},
isbn = {978-1-939133-25-0},
publisher = {USENIX Association},
address = {USA},
booktitle = {Proceedings of the Seventeenth Symposium on Usable Privacy and Security},
articleno = {26},
numpages = {20},
series = {SOUPS'21}
}

@inproceedings{brunken2023properly,
author = {Brunken, Lina and Buckmann, Annalina and Hielscher, Jonas and Sasse, M. Angela},
title = {"To do this properly, you need more resources": the hidden costs of introducing simulated phishing campaigns},
year = {2023},
isbn = {978-1-939133-37-3},
publisher = {USENIX Association},
address = {USA},
booktitle = {Proceedings of the 32nd USENIX Conference on Security Symposium},
articleno = {230},
numpages = {18},
location = {Anaheim, CA, USA},
series = {SEC '23}
}

@article{tanczer2021feel,
  title={‘i feel like we’re really behind the game’: perspectives of the united kingdom’s intimate partner violence support sector on the rise of technology-facilitated abuse},
  author={Tanczer, Leonie Maria and L{\'o}pez-Neira, Isabel and Parkin, Simon},
  journal={Journal of gender-based violence},
  volume={5},
  number={3},
  pages={431--450},
  year={2021},
  publisher={Policy Press}
}

@book{charmaz2006constructing,
  title={Constructing grounded theory: A practical guide through qualitative analysis},
  author={Charmaz, Kathy},
  year={2006},
  address={London, UK},
  publisher={sage}
}

@article{naeem2023step,
  title={A step-by-step process of thematic analysis to develop a conceptual model in qualitative research},
  author={Naeem, Muhammad and Ozuem, Wilson and Howell, Kerry and Ranfagni, Silvia},
  journal={International journal of qualitative methods},
  volume={22},
  pages={16094069231205789},
  year={2023},
  publisher={SAGE Publications Sage CA: Los Angeles, CA}
}

@article{ecabert2024implications,
  title={Implications of cyber incident reporting obligations on multinational organizations headquartered in Switzerland},
  author={Ecabert, Thomas and Muhly, Fabian and Zimmermann, Verena},
  journal={International Cybersecurity Law Review},
  volume={5},
  number={4},
  pages={585--614},
  year={2024},
  publisher={Springer}
}

@article{baker2007emotional,
  title={Emotional approach and problem-focused coping: A comparison of potentially adaptive strategies},
  author={Baker, John P and Berenbaum, Howard},
  journal={Cognition and emotion},
  volume={21},
  number={1},
  pages={95--118},
  year={2007},
  publisher={Taylor \& Francis}
}

@article{ghafur2019challenges,
  title={The challenges of cybersecurity in health care: the UK National Health Service as a case study},
  author={Ghafur, Saira and Grass, Emilia and Jennings, Nick R and Darzi, Ara},
  journal={The Lancet Digital Health},
  volume={1},
  number={1},
  pages={e10--e12},
  year={2019},
  publisher={Elsevier}
}

@article{karyda2005information,
  title={Information systems security policies: a contextual perspective},
  author={Karyda, Maria and Kiountouzis, Evangelos and Kokolakis, Spyros},
  journal={Computers \& security},
  volume={24},
  number={3},
  pages={246--260},
  year={2005},
  publisher={Elsevier}
}

@article{pfleeger2014weakest,
  title={From weakest link to security hero: Transforming staff security behavior},
  author={Pfleeger, Shari Lawrence and Sasse, M Angela and Furnham, Adrian},
  journal={Journal of Homeland Security and Emergency Management},
  volume={11},
  number={4},
  pages={489--510},
  year={2014},
  publisher={De Gruyter}
}

@inproceedings{lain2022phishing,
  title={Phishing in organizations: Findings from a large-scale and long-term study},
  author={Lain, Daniele and Kostiainen, Kari and {\v{C}}apkun, Srdjan},
  booktitle={2022 IEEE Symposium on Security and Privacy (SP)},
  pages={842--859},
  year={2022},
  address={San Francisco, CA, USA},
  publisher={IEEE}
}

@inproceedings{chen2025beyond,
author = {Chen, Xiaowei and Sch\"{o}ni, Lorin and Distler, Verena and Zimmermann, Verena},
title = {Beyond Deterrence: A Systematic Review of the Role of Autonomous Motivation in Organizational Security Behavior Studies},
year = {2025},
isbn = {9798400713941},
publisher = {Association for Computing Machinery},
address = {New York, NY, USA},
url = {https://doi.org/10.1145/3706598.3713122},
doi = {10.1145/3706598.3713122},
booktitle = {Proceedings of the 2025 CHI Conference on Human Factors in Computing Systems},
articleno = {919},
numpages = {28},
location = {Yokohama, Japan},
series = {CHI '25}
}

@article{freed2019my,
  title={" Is My Phone Hacked?" Analyzing Clinical Computer Security Interventions With Survivors of Intimate Partner Violence},
  author={Freed, Diana and Havron, Sam and Tseng, Emily and Gallardo, Andrea and Chatterjee, Rahul and Ristenpart, Thomas and Dell, Nicola},
  journal={Proceedings of the ACM on Human-Computer Interaction},
  volume={3},
  number={CSCW},
  pages={1--24},
  year={2019},
  publisher={ACM New York, NY, USA}
}

@article{beals2015framework,
  title={Framework for a taxonomy of fraud},
  author={Beals, Michaela and DeLiema, Marguerite and Deevy, Martha},
  journal={Financial Fraud Research Center},
  pages={39},
  year={2015}
}

@inproceedings{wunder2025achieving,
author = {Wunder, Julia and Wash, Rick and Renaud, Karen and Oliveira, Daniela A and Benenson, Zinaida},
title = {Achieving Resilience: Data Loss and Recovery on Devices for Personal Use in Three Countries},
year = {2025},
isbn = {9798400713941},
publisher = {Association for Computing Machinery},
address = {New York, NY, USA},
url = {https://doi-org.proxy.bnl.lu/10.1145/3706598.3714202},
doi = {10.1145/3706598.3714202},
booktitle = {Proceedings of the 2025 CHI Conference on Human Factors in Computing Systems},
articleno = {830},
numpages = {26},
location = {Yokohama, Japan},
series = {CHI '25}
}

@inproceedings{von2024beyond,
author = {Von Preuschen, Alexandra and Schuhmacher, Monika C. and Zimmermann, Verena},
title = {Beyond fear and frustration - towards a holistic understanding of emotions in cybersecurity},
year = {2024},
isbn = {978-1-939133-42-7},
publisher = {USENIX Association},
address = {USA},
booktitle = {Proceedings of the Twentieth Symposium on Usable Privacy and Security},
articleno = {33},
numpages = {20},
location = {Philadelphia, PA, USA},
series = {SOUPS '24}
}

@article{alhidaifi2024survey,
  title={A survey on cyber resilience: Key strategies, research challenges, and future directions},
  author={Alhidaifi, Saleh Mohamed and Asghar, Muhammad Rizwan and Ansari, Imran Shafique},
  journal={ACM computing surveys},
  volume={56},
  number={8},
  pages={1--48},
  year={2024},
  publisher={ACM New York, NY}
}

@misc{fbi2024,
  title        = {Internet Crime Report 2024},
  author       = {FBI},
  pages        = {47},
  year         = {2025},
  howpublished = {\url{https://www.ic3.gov/AnnualReport/Reports/2024_IC3Report.pdf}},
  note         = {Accessed: 2025-05-05}
}

@misc{eu_services,
  title        = {Payment services deal: More protection from online fraud and hidden fees},
  author       = {{European Parliament}},
  year         = {2025},
  url          = {https://www.europarl.europa.eu/news/en/press-room/20251121IPR31540/payment-services-deal-more-protection-from-online-fraud-and-hidden-fees},
  note         = {Accessed: 2025-11-28}
}

@article{vishwanath2020cyber,
  title={Cyber hygiene: The concept, its measure, and its initial tests},
  author={Vishwanath, Arun and Neo, Loo Seng and Goh, Pamela and Lee, Seyoung and Khader, Majeed and Ong, Gabriel and Chin, Jeffery},
  journal={Decision Support Systems},
  volume={128},
  pages={113160},
  year={2020},
  publisher={Elsevier}
}

@inproceedings{scamReddit,
  author    = {Elijah Bouma{-}Sims and Mandy Lanyon and Lorrie Faith Cranor},
  title     = {“Is this a scam?”: The Nature and Quality of Reddit Discussion about Scams},
  booktitle = {Proceedings of the 2025 ACM SIGSAC Conference on Computer and Communications Security (CCS '25)},
  year      = {2025},
  pages     = {1--15},
  address   = {Taipei, Taiwan},
  publisher = {ACM},
  doi       = {10.1145/3719027.3765030}
}

@inproceedings{oak2025hello,
  title={``Hello, is this Anna?'': Unpacking the Lifecycle of Pig-Butchering Scams},
  author={Oak, Rajvardhan and Shafiq, Zubair},
  booktitle={Twenty-First Symposium on Usable Privacy and Security (SOUPS 2025)},
  pages={1--18},
  publisher={USENIX Association},
  address={Seattle, WA, USA},
  year={2025}
}

@inproceedings{howe2012psychology,
author = {Howe, Adele E. and Ray, Indrajit and Roberts, Mark and Urbanska, Malgorzata and Byrne, Zinta},
title = {The Psychology of Security for the Home Computer User},
year = {2012},
isbn = {9780769546810},
publisher = {IEEE Computer Society},
address = {USA},
url = {https://doi-org.proxy.bnl.lu/10.1109/SP.2012.23},
doi = {10.1109/SP.2012.23},
booktitle = {Proceedings of the 2012 IEEE Symposium on Security and Privacy},
pages = {209–223},
numpages = {15},
series = {SP '12}
}

@inproceedings{deng2025auntie,
author = {Deng, Yue and He, Changyang and Zou, Yixin and Li, Bo},
title = {"Auntie, Please Don't Fall for Those Smooth Talkers": How Chinese Younger Family Members Safeguard Seniors from Online Fraud},
year = {2025},
isbn = {9798400713941},
publisher = {Association for Computing Machinery},
address = {New York, NY, USA},
url = {https://doi-org.proxy.bnl.lu/10.1145/3706598.3714137},
doi = {10.1145/3706598.3714137},
booktitle = {Proceedings of the 2025 CHI Conference on Human Factors in Computing Systems},
articleno = {864},
numpages = {17},
location = {Yokohama, Japan},
series = {CHI '25}
}

\appendix

\section{Recruiting questionnaire, informed consent, and interview protocol for the study}
\label{protocol}

\subsection{Recruiting questionnaire distributed on Prolific}
\textbf{Title:} Share your experience with cybercrimes

Welcome to participate in our research project, which aims to understand individuals' experience with various types of cybercrime. The findings of this study inform future mitigation strategies and better support mechanisms for cybercrime victims. We are a team of researchers from MPI-SP and the University of Luxembourg. 

This initial survey takes approximately 1-2 minutes to complete. It collects anonymous demographic information and asks whether you have experienced a cybercrime. You may also indicate your interest in participating in a follow-up interview study sharing your story in more depth. You are free to discontinue your participation in the questionnaire or interview at any point, without the need to provide any explanation. The interview lasts around 30-40 minute, and we compensate your time with £25 Prolific bonus (alternative: €30 gift voucher).

This study has been reviewed and approved by the MPI-SP Ethics Review Board and complies with the GDPR data processing. All information collected will be handled with the highest confidentiality. Survey responses are anonymous, and any identifying information collected for interview recruitment will be removed before publishing results.

If you have any questions about the study, please feel free to contact the research team: xiaowei.chen@mpi-sp.org

Please indicate that you understand the information provided and agree to participate in this survey (Selected choice).

\begin{itemize}
    \item No. I do not want to participate in the study.
    \item Yes. I want to participate in the study.
\end{itemize}

\begin{enumerate}[leftmargin=*]

\item What is your Prolific ID?

\item What is your gender? (Selected Choice) 

-- Woman; Man; Non-binary/third gender; Prefer not to say; Prefer to self-describe -- Text

\item Please indicate your age.

\item What is your current occupation?

\item What is your highest achieved degree? (Selected choice) 

-- High school diploma; Vocational training; Bachelor's degree; Master's degree; Doctoral degree; Prefer to self-describe -- Text

\item Have you experienced a cybercrime before? (Selected choice) 

-- No; Yes. Phishing/spoofing; Yes. Financial fraud; Yes. Cryptocurrency scam; Yes. Romance scam; Yes. Malware; Yes. Ransomware; Yes. Other cybercrimes (describe in text box) -- Text

\item Could you provide a brief account of what has happened?

\item When did the above-mentioned cybercrime happen? (fill in month/year, e.g., 06/2024)

\item People have different preferences when it comes to talking about cybercrime experiences. Which of the following statement best describes you? (Selected choice)  

-- I generally find it helpful to share with other of my experiences. I find it distressing to talk about and revisit my experiences. I am not sure; it depends on the situation.

\item Do you want to participate in our interview on cybercrimes? It will be conducted via Zoom remotely and last approximately 30 to 45 minutes, and you will receive a £25 Prolific bonus (alternative: €30 gift voucher) as a token of appreciation for your time.

\item Please indicate your availability by providing a preferred date and time (e.g., 4 July, 10:00 am).

\item Please indicate an email address that we can invite you for the interview via ZOOM.

\end{enumerate}

\subsection{Informed consent}
\label{consent}
(In addition to the consent form sent to each interviewee prior to the interview, we obtained informed consent before recording each session. )
[Welcome the participants] Briefly introduce the interviewer, a researcher at the University of Luxembourg (visiting MPI-SP), working on supporting people recover from cybercrimes. This is a collaborative project with researchers at MPI-SP. The interview will have three main parts:
\begin{itemize}
    \item First, we would like to know some details of the cybercrime you encountered (when, how, which platform, and your responses).
    \item Then, we want to ask how you coped with the incident and aspects that contributed to your recovery.
    \item Last, we are curious about your reflections and recommendations to others regarding similar cybercrimes.
\end{itemize}

The interview will last around 30 to 40 minutes. And we will compensate you with a £25 Prolific bonus (or a 30 euro gift Voucher) for your time contributing to this research. 

\textbf{The data we collect}: The interview will be audio recorded to allow us to transcribe the conversation with MAXQDA, a GDPR compliance transcription service. After transcription, the transcripts will be anonymized. Any identifiable information will be removed prior to publication. Your audio recordings will be deleted permanently after publication. During the interview, you can skip questions that you do not want to respond, and you can stop whenever you want. You also have the right to withdraw from the research after the interview. 

Do you have any questions regarding the interview or data collection \dots OK, now I will start the recording, and could you confirm that you give consent to be audio recorded as part of this study?

\subsection{Interview protocol}

\textbf{Section 1: Elicit a detailed account of the incident} (Some of these questions were inspired by Veisi et al.~\cite{veisi2025user}.)

\begin{enumerate}
    \item First, can you tell us your story of the cybercrime? Feel free to share as much as you like. This might be difficult to talk about, and you can stop whenever you want. 
    
    [Follow-up question, if they are not mentioned:] Which platform or application were you using when the incident occurred?
    
    [Follow-up question, if they are not mentioned:] What was the impact of the incident, for example, has the incident caused you some losses?
    
    \item Before this incident, have you experienced other cybercrimes targeting you? (How about your friends or family members?)
    \item How confident are you in managing your digital devices and online accounts?  
    \item Where did you learn these practices you just mentioned?   
    \item Have you had any forms of training related to cybersecurity or online safety? 
    \item How did you feel after the incident?  
    
[Follow-up question: Can you tell me a bit more about [the topic]? ]  
    
\end{enumerate}

\textbf{Section 2: Individual’s recovery process }

\begin{enumerate}
    \item To what extent have you recovered from the incident? [Allow the interviewee to define what recovery means to them.]
    \item How did you try to resolve the issue, if any?  
    \item Which aspects of your life experience supported your recovery?

[Follow-up question, if they are not mentioned:

        • Have you sought help from family, friends, colleagues, or online forums? 
        
        • Have you contacted the platform, device OS, or the application for help? If yes, how was the process? 
        
        • How was your impression of [the platform]? Do you think they would be interested in countering cybercrime?
        
        • Have you contacted Law enforcement, financial institutions, or insurance companies? (If not: How was your impression of the law enforcement? Do you think they would be interested in the cybercrime?)

\end{enumerate}
 
\textbf{Section 3: Lessons learned}

\begin{enumerate}
    \item Have you developed any new practices after the incident
    \item Have you made any changes to your online habits due to this incident?
    \item How do you currently protect yourself against similar crimes in light of this cyber incident?
    \item What advice would you offer to others who might fall into this incident based on your experience?
    \item Thank you so much for sharing your experience. Could you take a moment to reflect and summarize the key aspects that supported your recovery from the incident? 
    
\end{enumerate}

Debrief session: Address any questions the interviewees may have. Acknowledge that recovering from the experience of cybercrime may take time; and there are local support organizations that can provide assistance. Offer practical suggestions on how to protect themselves against the type of cybercrime they experienced.

\section{Cybercrime experience shared by study participants}
\label{scams}
\paragraph{P1 is a sales manager.} He travels regularly to meet clients in European and Asian countries. In 2018, when he just arrived in Malaysia for a business trip, he discovered around £2,500 had already been transferred out of his account. He realized that his bank account had been ``hacked.'' P1 contacted his bank immediately. Customer service transferred him to the cybercrime unit, reassuring him that everything would be under control. They helped him retrieve the money within two days. He never figured out how his \textbf{bank account was compromised}. Another cyber incident occurred in July 2025. P1 received two emails from \textbf{Qantas Airlines} (Australia) stating that his name, email, and phone number had been leaked due to a recent \textbf{cyber incident} on 30 June 2025. He should ``remain alert, especially through email, text messages, or telephone calls, particularly where the sender or caller purports to be from Qantas'' (Quote from the email content shared by P1).

\paragraph{P2 works as a funeral director at a company with hundreds of employees.} One day in 2025, while driving to work, he received a stream of notifications about payments made to \textbf{Uber Eats}. The deductions ranged between £20 and £30, totaling around £300. Amid the distractions caused by the continuous notifications of \textbf{unauthorized payments}, P2 had to find an opportunity to pull over on the motorway. He contacted the bank immediately, and they helped him block his card and recover the money. Prior to this incident, P2's \textbf{eBay account} had been \textbf{compromised} a couple of years earlier, and someone was able to place a random order using his account. P2 contacted PayPal, and the order was successfully canceled.

\paragraph{P3 works as a support worker.} One day in 2024, she received a message on \textbf{WhatsApp}: ``Are you interested in a part-time job?'' The person persuaded her to open a \textbf{crypto account} and invited her to create a \textbf{task account} on a website. After a few days, she ``earned'' nearly £400 through a couple of ``lucky orders,'' but to withdraw this money, she had to deposit some cryptocurrency into her account as a guarantee. After she deposited £500, her task account suddenly went into the negative. The person told her that another ``lucky order'' had just arrived. This time, she would get thousands after finishing the task. Luckily, P3's friend alerted her that this was a \textbf{task scam} after learning her story and advised her to block the scammer. P3 did not lose any additional money; however, she was unable to recover the £500 she had deposited into her task account. 

\paragraph{P4 hosts career counseling workshops for different UK schools.} In 2025, she received multiple alerts from her bank about failed payment attempts at various food shops. Although the transactions didn’t go through because she canceled the card a few days ago, it became clear that someone had somehow accessed her card details and tried to initiate \textbf{unauthorized payments}. A few years ago, P4 accidentally clicked on an ad link while using a free music converter website. That link turned out to be malicious, and a \textbf{malware} infected her laptop. P4 had to get help from her brother to remove it by deleting files and reinstalling the device's operating system.

\paragraph{P5 works as an account clerk.} He joined \textbf{a dating site} in 2022 and exchanged messages with a person who claimed to work abroad for the Foreign Office. The conversation seemed genuine at first, but soon the person started to request financial support for different reasons via \textbf{Instagram}, such as medical treatment and processing fees for gold coin legacy. Despite initial doubts, P5 ended up making several transfers to the person’s account. Eventually, the bank noticed the suspicious transaction and warned P5 that he was \textbf{being scammed}. In total, P5 had sent around £16,000. Fortunately, half of the amount was retrieved by his bank.

\paragraph{P6 works in risk modeling for financial institutions.} In December 2021, influenced by university peers and out of personal interest, she subscribed to investment advice from a \textbf{crypto coach} for £500. Some recommendations from this coach seemed legitimate, others dubious. One recommendation led her to download a beta cryptocurrency game, after which suspicious folders appeared on the desktop and Windows security was disabled. University IT support confirmed the presence of \textbf{Trojans} and removed the malware. After that, she also could not unsubscribe from the crypto coach’s service, and she eventually asked her bank to block payments.

\paragraph{P7 is a tram driver.} He experienced a targeted \textbf{bank service scam} in 2023. The day after P7 transferred £25,000 to pay off a mortgage, he received a call from someone claiming to be from the receiving bank. The caller knew P7’s name, bank, and transfer details, which made the call seem plausible at first. They told P7 the money hadn’t arrived and instructed him to initiate a second transfer. However, inconsistencies in the conversation and pressure to act quickly raised red flags. Eventually, P7 ended the call and checked with his bank about the transaction. P7 suspects his bank information might have been leaked through his phone or computer, though the exact source remains unclear.

\paragraph{P8 is self-employed.} He was in a desperate situation to earn money for a family member’s medical treatment in early 2024. Acting on advice from a media outlet, he invested £3,000 in the presale of a meme coin project. The project proved to be a \textbf{crypto rug pull}: the team behind it drained all raised funds before other investors could sell their tokens. P8 was unable to recover his losses. Through investigation with other investors in a Telegram group, they concluded that the presale was a scam, facilitated by exploitable flaws in the investment contract.

\paragraph{P9 works as a service manager.} In December 2024, when P9 was out with friends, he received a \textbf{phone call} from a person claiming to be from \textbf{Coinbase customer support}. A couple of days ago, P9 indeed contacted Coinbase via email for a transaction. The scammer deceived P9 into providing personal login information, which led to the loss of £200 worth of cryptocurrency from his Coinbase wallet. Despite contacting Coinbase and receiving standard responses, P9 could not recover the lost cryptocurrencies.

\paragraph{P10 is a care assistant.} In 2018, she sold her iPad (128GB) on eBay for €200. However, the buyer complained that the device was not as described and demanded a refund. P9 had little experience with eBay and felt intimidated by the threat of being reported to \textbf{eBay}, and agreed to reverse the transaction. Upon receiving the returned iPad, she discovered that the original \textbf{iPad had been replaced} with a 32GB storage one. After a series of legal processes, she still could not retrieve her original iPad when we interviewed her.

\paragraph{P11 is a researcher in psychology.} One day in 2018, she received a very angry email from a stranger, warning her to stop sending scam emails. When P11 read the thread, she discovered that the scam emails this person referred to had indeed been sent from her account. P11 logged into her email and noticed that some emails appeared and then disappeared immediately. Her \textbf{email account was compromised} and exploited by scammers. With technical support from her email provider, a German telecommunications company, P11 changed her passwords and even reset her client numbers at the company.

\paragraph{P12 is a researcher in cybersecurity.} In 2024, P12 moved to Germany for work. He was waiting for a delayed parcel when he received a fake \textbf{DHL delivery SMS}. Although he usually ignored such messages, this one was in German, and the timing made it seem plausible. In a rush, P12 clicked the link and entered his card payment details on a fraudulent DHL website to ``track'' the delivery. Within a minute, P12 realized it was a scam and blocked his card. Fortunately, he submitted the virtual card, which could be canceled conveniently. 

\paragraph{P13 works as a strategic advisor.} In 2023, one Saturday morning, she first received several suspicious SMS messages, followed by a phone call \textbf{impersonating her bank}. The scammer claimed there were issues with her account and even provided accurate financial details about her balances, which made the call appear credible. They asked her to log in to her account via mobile, but she refused, citing technical issues and preferring to use a computer. Eventually, P13 used another phone to contact her bank’s customer service directly. The staff confirmed that her account had been targeted and helped her lock the account. Later, P13 reckoned that her information had likely been leaked because her father’s \textbf{email was phished}, leading to scam calls to other family members as well. Although she suffered no financial loss, the incident made her more wary of online security.

\paragraph{P14 is a student in a master’s program in communication.} In 2024, while P14 was waiting for a parcel, she received an SMS that led to a \textbf{fraudulent delivery website} requesting personal and card information. Afterwards, reminded by her boyfriend, she blocked her bank card. Later, she received a phone call claiming to be her \textbf{bank customer service}, offering support to address suspicious account activity. The scammer then added multiple beneficiaries to her bank account and made several transfers, totaling €3,000. With support from her boyfriend and a family member, she blocked her account about two hours later after several transfers within her bank service team. P14 retrieved her loss after one week.

\paragraph{P15 is a researcher in cryptography.} In 2023, P15 experienced a \textbf{WhatsApp phishing} incident. Weeks after losing a family member, she was in a period of emotional distress. She received a forwarded message from a friend promoting a fake British Airways Black Friday offer on WhatsApp. P15 clicked the link and followed the instructions to forward the message to others before realizing it was a scam. Although she quickly stopped and took remedial actions, such as scanning her device, changing SIM cards, and restoring default settings, her phone number and email were exposed, and the password to an important file was accidentally deleted. Over the following months, P15 began receiving spam calls, phishing emails, and inappropriate messages on applications like Skype.

\paragraph{P16 is a freelance musical teacher.} When preparing for an exchange semester to Prague in 2023, P16 experienced a \textbf{rental scam} while searching for housing online. After receiving a quick and overly eager response from a ``landlord,'' she was pressured to pay immediately to secure the room. Feeling stressed and short on time, she transferred the money, only to later realize she was communicating with a scammer. Communication with her bank and the police was difficult, as the scam involved bank accounts in two EU countries. With help from her father, she continued to follow up, and very fortunately, the money was returned six months later without explanation from the bank.

\paragraph{P17 works as a material manager at a tech company.} In 2021, P17 experienced a \textbf{Facebook Messenger phishing}, where she clicked on a link, and automatic messages were sent from her account to all her contacts. After P17 had identified the phishing incident, she changed her password and tried to inform her contacts as quickly as possible. However, the scam spread rapidly through her network; some of her friends clicked the link as well and lost access to their accounts. Another incident happened in 2024. P17 received a \textbf{Facebook} message from her colleague indicating a money transfer. It turned out that the colleague's \textbf{account was breached} and exploited by attackers.

\paragraph{P18 works as a computer science professor.} He was contacted by an intended buyer via WhatsApp while selling an item on Blocket.se. The scammer sent him a link to a \textbf{fraudulent Blocket site}, requesting P18’s bank details to finalize the transaction. After verification with BankID, P18 found that 19,942 SEK had been transferred from his account and realized it was a fraud; he called his bank immediately, blocked his card, and reported the case to the police. He also attempted to contact TransferGo, the company handling the transfer to the scammer, but they were closed on Sunday; the next day, TransferGo confirmed nothing could be done as the money had already been sent to the recipient.


\end{document}